\documentclass[aps, prx, superscriptaddress, showpacs, twocolumn, reprint, longbibliography]{revtex4-1}

\usepackage{graphicx}
\usepackage{amssymb}
\usepackage{amsmath}
\usepackage{natbib}
\usepackage{braket}
\usepackage{makecell}

\newcommand{\fbar}{\bar{F}}

\newcommand{\eps}[1]{\varepsilon({\bf #1})}

\newcommand{\bfx}{{\bf x}}

\begin{document}

\title{Theory of reflectionless scattering modes}

\author{William R. Sweeney}
\email{william.sweeney@yale.edu}
\affiliation{Department of Physics, Yale University, New Haven, CT 06520, USA}

\author{Chia Wei Hsu}
\affiliation{Department of Applied Physics, Yale University, New Haven, CT 06520, USA}
\affiliation{Ming Hsieh Department of Electrical and Computer Engineering, University of Southern California, Los Angeles, California 90089, USA}

\author{A. Douglas Stone}
\affiliation{Department of Applied Physics, Yale University, New Haven, CT 06520, USA}
\affiliation{Yale Quantum Institute, Yale University, New Haven, CT 06520, USA}

\date{\today}

\begin{abstract}
We develop the theory of a special type of scattering state in which a set of asymptotic channels are chosen as inputs, and the complementary set as outputs, and there is zero reflection back into the input channels.
We show that in general an infinite number of such solutions exist at discrete frequencies in the complex $\omega$ (or energy) plane for any choice of the input/output sets.
Our results apply to linear electromagnetic and acoustic wave scattering and also to quantum scattering.
We refer to such states as reflection-zeros (R-zeros) when they occur off the real-frequency axis and as Reflectionless Scattering Modes (RSMs) when they are tuned to the real-frequency axis and exist as steady-state solutions.
Such reflectionless behavior requires a specific monochromatic input wavefront, given by the eigenvector of a filtered scattering matrix with eigenvalue zero.
RSMs may be realized either by tuning parameters of the scatterer which do not break flux-conservation (index tuning) or by adding gain or absorption (gain-loss tuning).
We show that in general only a single continuous parameter needs to be tuned to create an RSM for a given choice of input/output channels, and that such RSMs exist in all dimensions and for arbitrary geometries, including finite scatterers in free space. 
In addition, a coupled-mode analysis shows that RSMs are the result of a generalized type of critical coupling.
A symmetry analysis of R-zeros and RSMs implies that RSMs of flux-conserving cavities are bidirectional in the sense that input and output channels can be interchanged and the resulting state will also be an RSM at the same frequency. 
RSMs of cavities with gain and/or absorption are generically unidirectional and do not satisfy this interchange symmetry.  
Non-flux-conserving systems with ${\cal PT}$-symmetry have unidirectional R-zeros in complex-conjugate pairs, implying that for small enough ${\cal T}$-breaking their reflectionless states arise at real frequency, without the need of any parameter tuning; this explains the widely observed existence of unidirectional unit transmission resonances in one-dimensional ${\cal PT}$ systems, and their disappearance at high frequency as a result of spontaneous ${\cal PT}$-symmetry breaking.
A new type of exceptional point is shown to occur at this transition, leading to an observable change in the reflection lineshape.
Numerical examples of RSMs are given for one-dimensional cavities with and without gain/loss, a one-dimensional ${\cal PT}$ cavity, a two-dimensional multiwaveguide junction, and a two-dimensional deformed dielectric cavity in free space.
We outline and implement a general technique for solving such problems, which shows promise for designing photonic structures which are perfectly impedance-matched for specific inputs, or can perfectly convert inputs from one set of modes to a complementary set.
The theory also provides insight into the behavior of complex cavities and multiple-scattering systems.
\end{abstract}


\maketitle

\section{Introduction}

    \subsection{Resonances and Reflectionless Scattering Modes}

Resonant scattering of waves is a fundamental process in classical and quantum physics.
At certain input frequencies or energies, incident waves impinging on an object or inhomogeneous region (``the scatterer'') can excite normal modes of oscillation of the scatterer, leading to strong scattering and relatively long interaction times.
The normal modes of any such open system are resonances, alternatively called quasi-normal modes or Gamow states, which correspond to purely outgoing solutions of the wave equation with complex frequencies, $\omega = \omega_r -i\gamma$, where $\gamma = 1/2\tau > 0$, $\tau$ is the dwell time or intensity decay rate, and $Q=\omega_r \tau$ is the quality factor of the resonance~\cite{1928_Gamow_ZFP,1981_Bohm_JMP,1998_Ching_RMP,2018_Lalanne_LPR}.
In general, the resonance solutions are not physically realizable steady-states due to their exponential growth at infinity, but they determine the scattering behavior under steady-state (real frequency) harmonic excitation.
However in electromagnetic scattering with gain, the resonances {\it can} be realized physically and correspond to the onset of laser emission~\cite{1973_Lang_PRA,2014_Esterhazy_PRA}.
This threshold lasing state is a solution of the linear Maxwell's wave equation for a scattering geometry, but with no incident wave; it can only be realized at discrete frequencies and with the appropriate amount of gain to balance exactly scattering and absorption loss.

Relatively recently, another type of special electromagnetic state has been identified and studied: Maxwell solutions with only {\it incoming} waves at real frequencies, a phenomenon referred to as Coherent Perfect Absorption (CPA)~\cite{2010_Chong_PRL,2011_Wan_Science,2017_Baranov_NRM}.
This is the time-reversed solution of lasing at threshold, and corresponds to the incoming version of the lasing mode at the same frequency, but incident on a scatterer (or cavity) with a dielectric function $\varepsilon^* (\bf r)$, in which the spatial distribution of gain is replaced with absorption.
Importantly, the absorption is perfect only for the specific incident wavefront; generic inputs like a plane wave or a gaussian beam can still experience significant scattering.
Unlike in lasing, for which the system self-organizes to find the lasing mode once sufficient gain has been introduced, in CPA the input wave must be synthesized correctly, even if the susceptibility has been tuned to the appropriate amount of absorption.
Nevertheless, CPA demonstrates that engineering the absorption and input correctly allows the realization of a special scattering state with only incoming waves.  

Another example of gain-loss engineering, sometimes called non-hermitian engineering, are structures with symmetric absorption and gain distributions (i.e., parity-time symmetric, ${\cal PT}$), which has gotten much recent attention~\cite{1998_Bender_PRL,2012_Regensburger_Nature,2014_Feng_Science,2014_Hodaei_Science,2014_Peng_nphys,2017_Feng_nphoton,2018_ElGanainy_nphys,2019_Ozdemir_nmat,2019_Miri_Science}.
One general property of one-dimensional (1D) open ${\cal PT}$ structures is the existence of one-way reflectionless states~\cite{2011_Lin_PRL,2011_Hernandez-Coronado_PLA,2011_Longhi_JPA,2012_Ge_PRA,2012_Jones_JPA,2013_Feng_nmat,2014_Ramezani_PRL_2,2014_Midya_PRA,2015_Fleury_ncomms,2016_Rivolta_PRA,2016_Jin_SR,2016_Chen_OE,2016_Yang_OE,2017_Sarisaman_PRA,2018_Sarisaman_PRB}, summarized in the review~\cite{2017_Huang_nanoph}, where reflection is zero from one incident direction but not from the other, and the transmission is reciprocal and equal to unity~\cite{2012_Ge_PRA}.
While these unidirectional reflectionless resonances are present in all ${\cal PT}$ structures, they apparently disappear at sufficiently high frequency for a fixed value of gain-loss~\cite{2011_Hernandez-Coronado_PLA,2011_Longhi_JPA,2012_Ge_PRA,2012_Jones_JPA,2013_Mostafazadeh_PRA}.
One-way reflectionless states have also been studied in non-flux-conserving systems that do not possess ${\cal PT}$ symmetry~\cite{2013_Mostafazadeh_PRA,2013_Castaldi_PRL,2014_Chen_OE,2014_Wu_PRL,2015_Horsley_NP,2015_Wu_PRA,2017_Gear_NJP},
including those supporting constant-intensity waves~\cite{2017_Makris_LSA,2019_Brandstotter_PRB}.

In the current work we show that the reflectionless states of these structures with gain/loss represent only one special case of a much more general phenomenon that is related to, but distinct from, resonances and CPA.
In any open system, we can choose a subset of asymptotic channels as the input channels; demanding no reflection back into those channels defines a set of reflectionless scattering states which we show exist at discrete complex frequencies that we refer to as {\it reflection zeros} (R-zeros).
Through parameter-tuning or by imposing symmetry, an R-zero can be moved to a real frequency to become a steady-state solution that we refer to as a {\it reflectionless scattering mode} (RSM).
The R-zero/RSM concept applies to any open system in all dimensions.
This includes systems which do not naturally divide into left and right asymptotic regions: any subset of asymptotic channels may be chosen as input with the complementary set being the output, even when the input-only and output-only channels spatially overlap.
We will show that a countably infinite number of R-zeros always exist, and in principle any desired subset can be found computationally, for each cavity or structure of interest.
Moreover, generically it takes the variation of only one parameter describing the cavity to make a chosen R-zero real, resulting in an RSM.
RSMs can be engineered in a cavity or scatterer larger than the relevant wavelength; they can have absorption or gain or be passive, and are not required to have any special symmetry. 
The left-to-right and right-to-left RSMs in 1D ${\cal PT}$ structures are special because they naturally exist at real frequencies due to symmetry, a property that our theory will describe.

R-zeros are similar to resonances in that they both occur at a discrete set of complex frequencies.
However, they differ in that they satisfy different boundary conditions: for a left-incident R-zero the solution is purely incoming in the left asymptotic region and purely outgoing in the right asymptotic region, while for resonances the solution is purely outgoing on both sides.
While the boundary condition of R-zeros differs from that of the resonances, it corresponds to the {\it same number of conditions} imposed infinitely far from the scatterer. 
In this sense, the existence of R-zeros is no more extraordinary than the existence of resonances.

A recent work~\cite{2018_Bonnet-BenDhia_PRSA}, closely related to the present one, studied the R-zeros of obstacles in a few-mode acoustic waveguide. 
There, the outgoing boundary condition on the right is achieved using a perfectly matched layer (PML), and the incoming boundary condition on the left is achieved using a complex-conjugated PML. The tuning of R-zeros to the real axis to realize steady-state solutions (RSMs) was not discussed.  Moreover, 
the PML-based approach requires the input channels to be spatially separated from the output channels, limiting its applicability in arbitrary scattering geometries, such as finite-sized scatterers in free space (where the asymptotic channels are angular-momentum states), or planar scatterers illuminated with a finite numerical aperture (where the input channels consist of a subset of incident angles), or multimode waveguides with input and output channels occurring in the same waveguide.
The theory presented in this work is more general, applicable to any open system and for any choice of input/output channels.

We will provide two methods to find R-zeros in such systems, involving either a filtered scattering matrix or the underlying wave operator, and derive an explicit connection between the two methods that also highlights the role of bound states in the continuum~\cite{2016_Hsu_NRM}.
In the wave-operator approach, incoming and outgoing boundary conditions for arbitrary choices of channels can be implemented through the boundary-matching methods~\cite{2005_prog_in_optics} (see section \ref{sec:BC_MM}).

An RSM with a single input channel is equivalent to ``critical coupling'' to a cavity~\cite{Haus_book,2000_Yariv,2000_Cai}, with appropriate caveats to be discussed below.
However the general RSM concept is intrinsically multichannel, and to our knowledge has not been explored previously in any generic context.
RSMs can be designed to achieve perfect impedance-matching of multichannel inputs or to achieve perfect mode-conversion, hence we believe that RSMs open up a very promising avenue for design of photonic structures.
Moreover, the RSM concept applies to {\it any wave scattering system}, for example sound waves, waves of cold atoms or condensates, and other quantum systems.
In fact, unlike lasing or CPA which are non-unitary by definition (i.e., not conserving energy flux), steady-state RSMs can be flux-conserving and do not require the availability of absorption or gain as a design resource.
We will discuss engineering of passive RSMs (without loss or gain) below.

    \subsection{The Scattering Matrix\label{sec:intro to S-matrix}}

To make the RSM concept more precise we must now introduce the starting point of the theory, the scattering matrix (S-matrix) of an open wave system. 
The system generally consists of an inhomogeneous scattering region, outside of which are asymptotic regions that extend to infinity.
To support resonance effects the scattering region needs to be larger than wavelength of the input waves within the scattering medium.
The asymptotic regions are assumed to be time-reversal invariant and to have some form of translational invariance, e.g., vacuum or uniform dielectric, or a finite set of waveguides, or an infinite periodic photonic crystal.
We also focus on media in which the scattering forces are short-range, so that the complications associated with long-range Coulomb or other forces are absent.
This simplification holds for electromagnetic waves and sound waves in almost all cases, and for the remainder of this work we will develop the theory in terms of classical Maxwell waves.
RSMs also exist as an exact phenomenon in quantum electrodynamics, and we will briefly comment on their manifestation there in the conclusions section.

A linear and static photonic structure is described by its dielectric function $\varepsilon({\bf r,\omega})$, which can be complex-valued, in which case its imaginary part describes absorption and/or gain. 
The assumed linearity allows the theory to concentrate on scattering at a single real frequency, $\omega$; time-dependent scattering can be studied by superposing solutions.
The translational symmetry of the asymptotic regions allows one to define $2N$ power-orthogonal propagating ``channel states'' at each $\omega$.
Based on the direction of their fluxes, the $2N$ channels can be unambiguously grouped into $N$ incoming and $N$ outgoing states, which are related by time-reversal.
Familiar examples of channels include the guided transverse modes of a waveguide and orbital angular-momentum waves in free space, with one channel per polarization.
In the waveguides, the finite number and width of the waveguides lead to a finite $N$ for a given $\omega$
, whereas for the case of a finite scatterer/cavity in free space the number of propagating angular-momentum channels is countably infinite.
However, a finite scatterer of linear scale $R$, with no long-range potential outside, will interact with only a finite number of angular momentum states, such that $l_{\rm max} \sim \sqrt{ \bar{\epsilon} }R \omega /c$, where $\bar{\epsilon}$ is the spatially-averaged dielectric function in the scattering region, and $c$ is the speed of light.
Hence for each $\omega$ we can reasonably truncate the infinite dimensional channel-space to a finite, $N$-dimensional subspace of relevant channels.

\begin{figure}
    \centering
    \includegraphics[width=0.95\columnwidth]{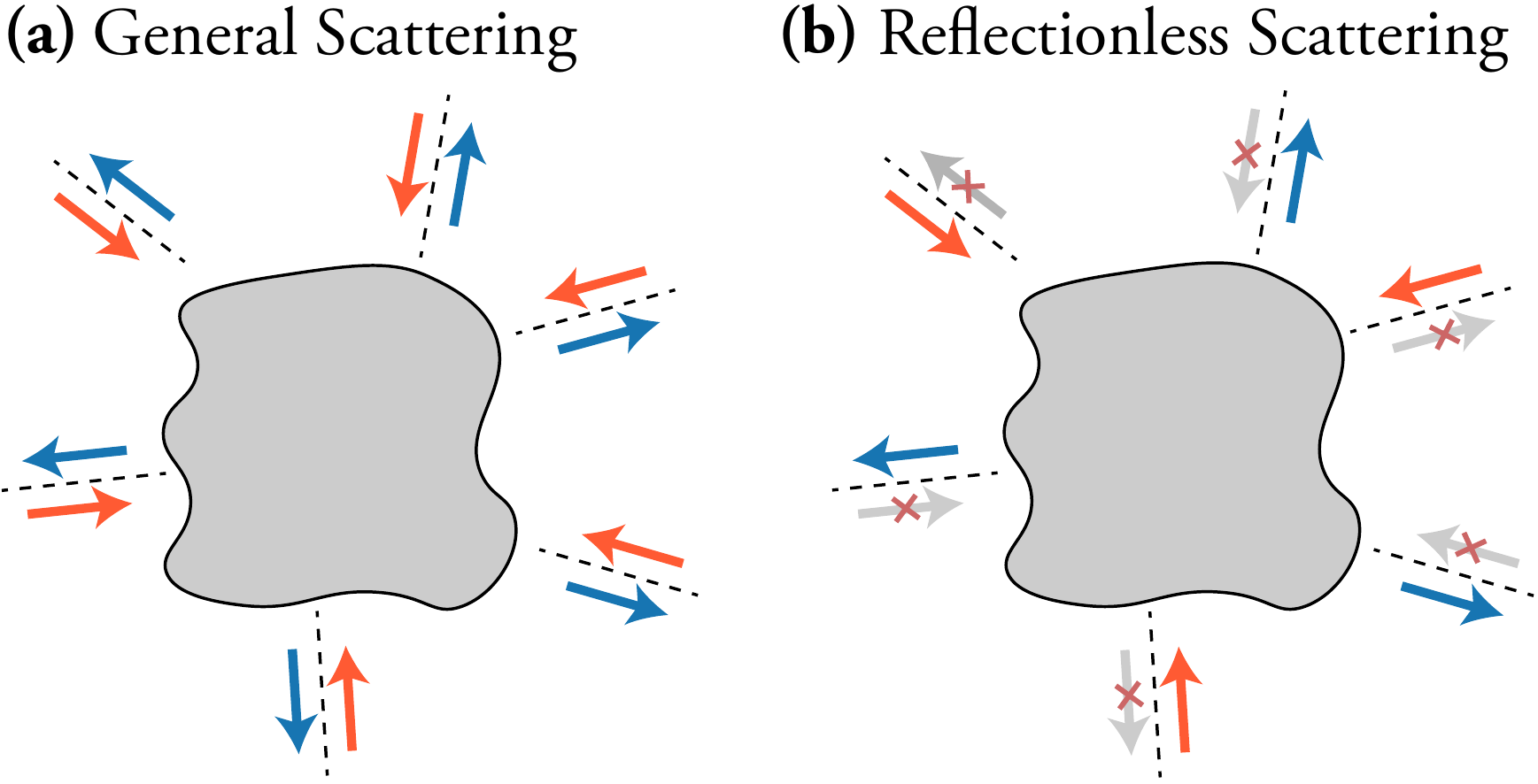}
    \caption{
    (Color) Schematic depicting a general scattering process (a) and reflectionless process (b).
    A finite scatterer/cavity interacts with a finite set of asymptotic incoming and outgoing channels, indicated by the red and blue arrows, respectively, related by time-reversal.
    These channels may be localized in space (e.g., waveguide channels) or in momentum space (e.g., angular-momentum channels).
    {\bf (a)} In the general case without symmetry, all incoming channels will scatter into all outgoing channels. 
    {\bf (b)} There exist reflectionless states, for which there is no reflection back into a chosen set of incoming channels (the inputs), which in general occur at discrete complex frequencies and do not correspond to a steady-state harmonic solution of the wave equation.
    However, with variation of the cavity parameters, a solution can be tuned to have a real frequency, giving rise to a steady-state reflectionless scattering process for a specific coherent input state, referred to as a Reflectionless Scattering Mode (RSM).
    }
    \label{fig:schematic}
\end{figure}

A general scattering process then consists of incident light, propagating along the $N$ incoming channels, interacting with the scatterer and then propagating out to infinity along the $N$ outgoing channels, as illustrated in Fig.~\ref{fig:schematic}(a).
The partial scattering of a single incoming channel into the outgoing channel which is its time-reverse defines a reflection coefficient, and scattering into each other channel defines the transmission coefficients.
In the channel basis, the wavefronts of the incoming and outgoing fields are given by length-$N$ column vectors ${\boldsymbol \alpha}$ and ${\boldsymbol \beta}$, normalized such that ${\boldsymbol \alpha}^\dagger {\boldsymbol \alpha}$ and ${\boldsymbol \beta}^\dagger {\boldsymbol \beta}$ are proportional to the total incoming and total outgoing energy flux, respectively.
The $N$-by-$N$ scattering matrix ${\bf S}(\omega)$, which relates ${\boldsymbol \alpha}$ and ${\boldsymbol \beta}$ at frequency $\omega$ is defined by
\begin{equation}
    \label{eq:S}
    {\boldsymbol \beta} = {\bf S}(\omega) {\boldsymbol \alpha},
\end{equation}
and comprises all of the reflection and transmission coefficients.
In reciprocal systems, the S-matrix is symmetric, ${\bf S} = {\bf S}^T$~\cite{2013_Jalas_nphoton}.
If the scatterer is lossless (i.e., $\varepsilon$ is real everywhere), then any incoming state leads to a non-zero flux-conserving output, and the S-matrix is unitary.

The S-matrix, being well-defined at all real frequencies (absent self-oscillating solutions, i.e., lasing), can be extended to complex frequencies via analytic continuation.

\section{General Theory of R-zeros/RSMs}

We will now define R-zeros and RSMs for a general geometry, prove the existence of solutions in the complex-frequency plane, and analyze the properties of R-zeros/RSMs based on the existence of different symmetries of the scattering system.  

    \subsection{R-zeros and RSMs\label{sec:R-zeros}}

{\it Reflection zeros} (R-zeros) are specific solutions of the scattering problem, typically at a complex frequency, for which there is no back-reflection into a given set of channels.
A {\it reflectionless scattering mode} (RSM) occurs when an R-zero is tuned to a real frequency, and is therefore a specially constrained steady-state solution of the more general R-zero problem.
To define an R-zero problem, we specify $N_{\rm in}$ of the incoming channels to be ``input channels'' which carry incident flux but no outgoing flux, and are thus reflectionless, with $0<N_{\rm in}<N$.
The complementary set of $N_{\rm out}=N-N_{\rm in}$ outgoing channels (the ``output channels'') carry any outbound flux. 
This is illustrated in Fig.~\ref{fig:schematic}(b).

Let ${\boldsymbol \alpha}_{\rm in}$ denote the $N_{\rm in}$-component vector containing the input amplitudes of such a reflectionless incident field, and ${\bf R}_{\rm in}(\omega)$ denote the square $N_{\rm in}$-by-$N_{\rm in}$ ``generalized reflection matrix'', which is the submatrix of the analytically continued ${\bf S}(\omega)$ defined by the specified input channels, i.e., $(R_{\rm in})_{i,j} = S_{n_i, n_j}$, where  $i,j=1,2,...,N_{\rm in}$ and $\{n_i\}$ is the set of $N_{\rm in}$ channels.
At the complex frequency $\omega_{\rm RZ}$ of an R-zero, the absence of reflection can be expressed as
\begin{equation}
\label{eq:RSM definition}
    {\bf R}_{\rm in}(\omega_{\rm RZ}){\boldsymbol \alpha}_{\rm in} = {\bf 0}, 
\end{equation}
which is the formal definition of an R-zero.
In other words, the R-zero frequencies are those at which ${\bf R}_{\rm in}(\omega)$ has a zero eigenvalue, the corresponding eigenvector ${\boldsymbol \alpha}_{\rm in}$ being the reflectionless incident wavefront. 
More generally, Eq.~\eqref{eq:RSM definition} defines a nonlinear eigenproblem where $\omega_{\rm RZ}$ is the eigenvalue, and methods for solving general nonlinear eigenproblems~\cite{2000_Golub_JCAM,2004_Voss_BIT,2010_Asakura_JIAM,2011_Su_SIAM,2012_Beyn_LA} may be used to solve it.

Eq.~\eqref{eq:RSM definition} can also be solved by finding the roots of the complex scalar function $\det {\bf R}_{\rm in}(\omega)$:
\begin{equation}
    \label{eq:det_Rin_zero}
    \det {\bf R}_{\rm in}(\omega_{\rm RZ}) = 0.
\end{equation}
Since the generalized reflection matrix can be computed for any open scattering system using standard methods, Eqs.~\eqref{eq:RSM definition}~\&~\eqref{eq:det_Rin_zero} provide universal recipes for solving the R-zero problem, for each of the $2^{N}-2$ choices of reflectionless input channels.

We will show in Section~\ref{sec:rigorous RSM} that $\det{\bf R}_{\rm in}(\omega)$ is a rational function of $\omega$.
Therefore, near each zero to leading order $\det{\bf R}_{\rm in}(\omega) \propto (\omega - \omega_{\rm RZ})$, so that the phase angle $\arg(\det {\bf R}_{\rm in})$ winds by $2\pi$ along a counterclockwise loop around each $\omega_{\rm RZ}$ in the complex-frequency plane.
This shows that each R-zero is a topological defect in the complex-frequency plane with a topological charge of $+1$~\cite{1979_Mermin_RMP}.
As such, R-zeros are robust: when the optical structure $\varepsilon({\bf r})$ is perturbed, $\omega_{\rm RZ}$ moves in the complex-frequency plane but cannot suddenly disappear, similar to topological defects in other systems~\cite{2014_Zhen_PRL,2017_Guo_PRL}.
For an R-zero to disappear, it must annihilate with another toplogical defect with charge $-1$, which we will show later is a pole of the S-matrix (i.e., resonance).
When parameters of the system are tuned such that $n>1$ R-zeros meet at the same frequency, they superpose and form a topological defect with charge $+n$ where $\det{\bf R}_{\rm in} \propto (\omega - \omega_{\rm RZ})^n$; we will show later that these are exceptional points (EPs) of a wave operator and provide an explicit example in Fig.~\ref{fig:PT}(b--e).

Even though the case of $N_{\rm in}=N$ leads to a well-defined ${\bf R}_{\rm in}={\bf S}$, we exclude it from the R-zero framework since the corresponding $\omega_{\rm RZ}$ is already understood as a zero of the S-matrix~\cite{2010_Chong_PRL}, and for flux-conserving systems is constrained to the upper half of the complex-frequency plane.
The case $N_{\rm in}=0$, corresponding to resonance, is also excluded, as here ${\bf R}_{\rm in}$ and the associated R-zeros are not defined.

The simultaneous absence of reflection in all input channels for the R-zero incident wavefront is due to interference: the reflection amplitude of each input channel $i$ destructively interferes with the interchannel scattering from all the other input channels, $(R_{\rm in})_{ii}{\alpha}_i+\sum_{j\neq i} (R_{\rm in})_{ij}{\alpha}_j=0$, which is precisely Eq.~\eqref{eq:RSM definition}.
The scattering (``transmission'') into the output channels is not obtained from solving this equation alone, and must be determined by solving the full scattering problem at $\omega_{\rm RZ}$.

While Eq.~\eqref{eq:RSM definition} or Eq.~\eqref{eq:det_Rin_zero} already provides a concrete recipe for solving R-zeros, we will further analyze $\det {\bf R}_{\rm in}(\omega)$ to shed light on the existence of R-zeros in the complex-frequency plane and to reveal their relation to the wave operator and to resonances. We do so in the next two sections.

\subsection{Wave-operator Representation of the S-matrix\label{sec:rigorous S}}

In this section, we introduce a wave-operator representation of ${\bf S}(\omega)$ and $\det {\bf S}(\omega)$, which we will adapt in the next section to analyze ${\bf R}_{\rm in}(\omega)$ and $\det {\bf R}_{\rm in}(\omega)$.

Consider a wave operator $\hat A(\omega)$ acting on state $\ket{\omega}$, which satisfies $\hat A(\omega)\ket{\omega}=0$.
For electromagnetic scattering, $\braket{{\bf r}|\omega}$ is the magnetic field, ${\bf H}({\bf r})$, under harmonic excitation at $\omega$, and the Maxwell operator at frequency $\omega$ is given by
\begin{equation}
\label{eq:Maxwell Operator}
\braket{{\bf r^\prime}|\hat A(\omega)|{\bf r}}=\delta({\bf r-r^\prime})\left\{\left(\frac{\omega}{c}\right)^2-\nabla\times\left(\frac{1}{\varepsilon({\bf r})}
\nabla\times\right)\right\}.
\end{equation}
Divide the system into two regions: the finite, inhomogeneous scattering region $\Omega$ in the interior, and the exterior asymptotic region $\bar\Omega$ that extends to infinity, which possesses a translational invariance broken only by the boundary, $\partial\Omega$, between $\Omega$ and $\bar\Omega$.
We separate operator $\hat A(\omega)$ into three pieces,
\begin{equation}
    \hat A(\omega) = \hat A_0(\omega)+\hat A_c(\omega)+\hat V(\omega),
\end{equation}
with $\hat A_0(\omega)$ identical to $\hat A(\omega)$ in region $\Omega$ and zero elsewhere, $\hat A_c(\omega)$ identical to $\hat A(\omega)$ in region $\bar{\Omega}$ and zero elsewhere, and the coupling term $\hat V(\omega)$ between the two regions defined via $\hat{V}(\omega) \equiv \hat{A}(\omega) - \hat{A_0}(\omega) - \hat{A_c}(\omega)$.

The auxiliary, closed-cavity wave operator $\hat A_0(\omega)$ on $\Omega$ admits a discrete spectrum of the form $\hat A_0(\omega_\mu)\ket{\mu}=0$ with eigenvalues $\{\omega_\mu\}$.
We explicitly do not assume $\hat A_0(\omega)$ to be hermitian, as one major focus of this work is to use absorption or gain to tune the R-zeros frequencies to become real, creating an RSM.
The matrix ${\bf A}_0(\omega)$ is naturally defined by its matrix elements
\begin{equation}
    \label{eq:A0}
    A_0(\omega)_{\mu \nu} = \braket{\mu|\hat A_0(\omega)|\nu}.
\end{equation}
The auxiliary asymptotic wave operator $\hat A_c(\omega)$ on $\bar\Omega$ has a continuous spectrum.
It generates the propagating channel modes via $\hat A_c(\omega)\ket{\omega,n}=0$ ($n$ here is an integer or set of integers which uniquely specify the asymptotic channels for each real frequency, $\omega$).
By construction, the only non-vanishing matrix elements of $\hat V$ are those between closed and continuum states, so that the matrix ${\bf W}(\omega)$, not necessarily square, is given by
\begin{equation}
    W(\omega)_{\mu n} = \braket{\mu|\hat V(\omega)|n,\omega},
\end{equation}
and contains all the information in $\hat V(\omega)$.

A general relation~\cite{1969_Mahaux_book,1997_Beenakker_RMP,2003_Viviescas_PRA,2017_Rotter_RMP} between the matrices ${\bf S}$, ${\bf A}_0$, and ${\bf W}$, originally developed in the study of nuclear reactions as the continuum-shell model or shell-model approach
~\cite{1957_Bloch,1958_Feshbach_ARNS,1958_Feshbach_AP,1961_Fano_PR,1966_Lane_PR,1969_Mahaux_book,1985_Nishioka_PLB, 2000_Dittes_PR, 2003_Sadreev_JPA}, is
\begin{equation}
\label{eq: Heidelberg}
    {\bf S}(\omega) = {\bf I}_N - 2 \pi i {\bf W}_p^\dagger(\omega) {\bf G}_{\rm eff}(\omega) {\bf W}_p(\omega),
\end{equation}
where ${\bf I}_N$ is the $N$-by$N$ identity, and the effective Green's function ${\bf G}_{\rm eff}(\omega)$ is the inverse of the effective wave operator ${\bf A}_{\rm eff}(\omega)$:
\begin{gather}
\begin{split}
    \label{eq:Aeff_and_Sigma}
    {\bf G}_{\rm eff} \equiv {\bf A}_{\rm eff}^{-1}, \qquad
    {\bf A}_{\rm eff} \equiv {\bf A}_0^\prime - {\boldsymbol \Sigma}^R,\\
    {\boldsymbol \Sigma}^R \equiv {\boldsymbol \Delta} - i{\boldsymbol \Gamma},\qquad
    {\boldsymbol \Gamma} \equiv \pi {\bf W}_p {\bf W}_p^\dagger,\\
    {\boldsymbol \Delta}(\omega)\equiv{\rm p.v.}\int d\omega^\prime \frac{{\bf W}_p(\omega^\prime) {\bf W}_p^\dagger(\omega^\prime)}{\omega^\prime-\omega}.
\end{split}
\end{gather}
For conciseness we have and will continue to suppress the dependence on $\omega$, except when needed for clarity.
The matrix ${\bf W}_p$ is the full coupling matrix ${\bf W}$ restricted to the $N$ propagating channels, while ${\bf A}_0^\prime$ is the closed-cavity wave operator ${\bf A}_0$ plus a hermitian modification that includes the effect of evanescent channel states.
The operator ${\boldsymbol \Sigma}^R$ is the self-energy, and acts only on the boundary $\partial\Omega$; ${\boldsymbol \Delta}$ and $-i{\boldsymbol \Gamma}$ are its hermitian and anti-hermitian components.

The resonances of the system are the eigenmodes of the non-hermitian, nonlinear eigenvalue problem $\hat{A}_{\rm eff}(\omega_p) \ket{\omega_p} = 0$, with eigenvalues $\{\omega_p\}$ that are the complex-valued resonance frequencies, which generally form a countably infinite set.
Note that the frequency dependence of the self-energy makes the eigenproblem nonlinear.
Since ${\boldsymbol \Gamma}$ is positive semidefinite, the self-energy due to openness generally contributes a negative imaginary part to the resonance frequencies, pushing the poles of ${\bf G}_{\rm eff}(\omega)$ and ${\bf S}(\omega)$ into the lower half of the complex-frequency plane, exactly capturing the effect of openness.

A series of manipulations applied to Eq.~\eqref{eq: Heidelberg} yields a powerful identity for $\det{\bf S}(\omega)$.
Using the ``push-through identity'' of linear algebra~\cite{Bernstein_Matrix_book}, we move the interaction matrix ${\bf W}_p$ across ${\bf G}_{\rm eff}$ to obtain ${\bf S} = ({\bf 1}+i{\bf K})/({\bf 1}-i{\bf K})$, where the reactance matrix ${\bf K} \equiv \pi {\bf W}_p^\dagger ({\bf A}_0^\prime-{\boldsymbol \Delta})^{-1} {\bf W}_p$~\cite{1967_MacDonald_PR,1982_Newton_book}.
Taking the determinant of both sides and dividing numerator and denominator by $\det ({\bf A}_0^\prime-{\boldsymbol \Delta})$, we arrive at
\begin{equation}
\label{eq:det(S)}
    \det {\bf S}(\omega) = \frac{\det{\boldsymbol (}{\bf A}_0^\prime(\omega)-{\boldsymbol \Delta}(\omega) - i{\boldsymbol \Gamma}{\boldsymbol )}}{\det{\boldsymbol (}{\bf A}_0^\prime(\omega)-{\boldsymbol \Delta}(\omega) + i {\boldsymbol \Gamma} 
    {\boldsymbol )}}.
\end{equation}
Note that Eq.~\eqref{eq:det(S)} is not a simple identity of linear algebra: 
the left-hand side is the standard determinant of an $N$-by-$N$ square matrix, while the right-hand side is a ratio of functional determinants of differential operators on an infinite-dimensional Hilbert space.
See Appendix~\ref{ap:detailed_derivation_S} for more detail on the derivation of Eq.~\eqref{eq:det(S)}.
A related expression was given in Refs.~\cite{2013_Grigoriev_PRA, 2019_Krasnok_arXiv}.

Eq.~\eqref{eq:det(S)} reveals that the analytically continued $\det {\bf S}(\omega)$ has zeros at a countably infinite set of complex frequencies $\{\omega_z\}$, which are the eigenvalues of the nonlinear eigenproblem $[\hat{A}_0(\omega_z)-\hat\Sigma^A(\omega_z)]\ket{\omega_z}=0$, where $\hat\Sigma^A = \hat \Delta + i\hat\Gamma$;
we refer to them as the {\it zeros of the S-matrix}.
CPA arises when, by tuning the degree of absorption in the system, one member of this set reaches a real frequency and becomes a steady-state solution.
As described in the introduction, the time-reverse of CPA is threshold lasing (purely outgoing solutions at real frequency).
We can now generalize that as follows: the time-reverse of the state corresponding to a complex zero of the S-matrix is a resonance.
For a lossless scatterer, time reversal symmetry implies that $\omega_z = \omega_p^*$, which is consistent with Eq.~\eqref{eq:det(S)} since ${\boldsymbol (}\hat\Sigma^A(\omega){\boldsymbol )}^\dagger = \hat\Sigma^R(\omega^*)$ when $\varepsilon({\bf r})$ is real.

While an eigenvalue of $\hat A_0(\omega)-\hat\Sigma^R(\omega)$ typically corresponds to a pole of $\det {\bf S}(\omega)$ and an eigenvalue of $\hat A_0(\omega)-\hat\Sigma^A(\omega)$ to a zero of $\det {\bf S}(\omega)$, there is one important exception.
Eq.~\eqref{eq:det(S)} shows that if $\omega_p = \omega_z$ is the simultaneous eigenvalue of both, $\det {\bf S}(\omega)$ may be neither zero nor infinite.
Such an exception happens at a bound state in the continuum (BIC)~\cite{2016_Hsu_NRM}, which contains neither incoming nor outgoing radiation and exists at a real frequency.
A BIC is invariant under time-reversal, and does not affect ${\bf S}(\omega)$ since it is decoupled from far-field radiation.
In the topological-defect picture, a BIC implies that a $+1$ charge (S-matrix zero) annihilates with an $-1$ charge (resonance) on the real axis.

    \subsection{Analytic Properties of R-zeros and RSMs\label{sec:rigorous RSM}}

We now adapt this formalism to treat R-zeros/RSMs, which can by analyzed in a similar fashion through the relation between the generalized reflection matrix ${\bf R}_{\rm in}$ and the S-matrix.
Let $F$ be the set of $N_{\rm in}$ filled input channels, which fixes $\fbar$, the set of $N_{\rm out}$ filled output channels (the remaining incoming and outgoing channels will carry no flux).
Because the channel ordering in ${\bf S}$ is arbitrary, for convenience we permute ${\bf S}$ such that the $N_{\rm in}$ input channels correspond to the upper-left block of ${\bf S}$, namely,
\begin{equation}
    \label{eq:Sin block form}
    {\bf S}(\omega) = \begin{pmatrix} 
        {\bf R}_{\rm in}(\omega) & {\bf T}_2(\omega) \\
        {\bf T}_1(\omega) & {\bf R}_{\rm out}(\omega)
    \end{pmatrix}.
\end{equation}
To extract ${\bf R}_{\rm in}$ from ${\bf S}$, let us define the filtering matrices ${\bf F}$ and $\bar{\bf F}$, where ${\bf F}:{\mathbb C}^N\rightarrow{\mathbb C}^{N_{\rm in}}$ reduces the dimension of the channel space from $N$ to $N_{\rm in}$, i.e., $F_{ij}=\delta_{ij}$ for $i\le N_{\rm in}$, $j\le N$, and $\bar{\bf F}$ reduces from $N$ to $N_{\rm out}$.
It follows that 
\begin{equation}
    \label{eq:F identites}
    {\bf FF}^\dagger = {\bf I}_{N_{\rm in}},\ 
    \bar{\bf F}\bar{\bf F}^\dagger= {\bf I}_{N_{\rm out}},\  
    {\bf F}^\dagger {\bf F}+\bar{\bf F}^\dagger \bar{\bf F}={\bf I}_N.
\end{equation}
With these definitions we have
\begin{equation}
    \label{eq:Rin_S}
    {\bf R}_{\rm in}(\omega) = {\bf FS}(\omega){\bf F}^\dagger.
\end{equation}
Using Eqs.~\eqref{eq: Heidelberg} and~\eqref{eq:Rin_S},
\begin{gather}
\begin{split}
    \label{eq:Rin first}
    {\bf R}_{\rm in} = {\bf I}_{N_{\rm in}} - 2 \pi i {\bf W}_F^\dagger {\bf G}_{\rm eff} {\bf W}_F,\\
    \quad {\bf W}_F \equiv {\bf W}_p {\bf F}^\dagger,
    \quad {\bf W}_{\fbar} \equiv {\bf W}_p\bar{\bf F}^\dagger.
\end{split}
\end{gather}
In analogy to what we did to derive the K-matrix representation of ${\bf S}$, we now push ${\bf W}_F$ through ${\bf G}_{\rm eff}$ to get ${\bf R}_{\rm in} = ({\bf I}_{N_{\rm in}}+i {\bf K}_F)/({\bf I}_{N_{\rm in}}-i {\bf K}_F)$, where ${\bf K}_F \equiv \pi {\bf W}_F^\dagger \bar{{\bf G}}_0 {\bf W}_F$, and $\bar{\bf G}_0 \equiv [{\bf A}_0^\prime - {\boldsymbol \Delta} + i{\boldsymbol \Gamma}_F]^{-1}$.
The self-energies restricted to the input channels is
\begin{gather}
    \label{eq:RSM self-energy}
\begin{split}
    {\boldsymbol \Sigma}^R_{F}(\omega) \equiv {\boldsymbol \Delta}_{F}(\omega) - i{\boldsymbol \Gamma}_{F}(\omega),\\
    {\boldsymbol \Delta}_{F}(\omega) \equiv {\rm p.v.}\int d\omega^\prime\, \frac{{\bf W}_{F}(\omega^\prime){\bf W}_{F}^\dagger(\omega^\prime)}{\omega^\prime-\omega},\\
    {\boldsymbol \Gamma}_{F}(\omega) \equiv \pi {\bf W}_{F}(\omega){\bf W}_{F}^\dagger(\omega),
    \end{split}
\end{gather}
and similarly for $\fbar$.
Taking the determinant of both sides of Eq.~\eqref{eq:Rin first} and applying the same procedure used to derive Eq.~\eqref{eq:det(S)}, we get
\begin{equation}
\label{eq:det(R)}
    \det {\bf R}_{\rm in}(\omega)
        = \frac{\det\boldsymbol{(}{\bf A}_0(\omega)-{\boldsymbol \Delta}(\omega)-i[{\boldsymbol \Gamma}_F(\omega) - {\boldsymbol \Gamma}_{\fbar}(\omega)]\boldsymbol{)}}{\det{\boldsymbol (}{\bf A}_0(\omega)-{\boldsymbol \Delta}(\omega)+i{\boldsymbol \Gamma}{\boldsymbol )}},
\end{equation}
where we have used Eq.~\eqref{eq:F identites} to write ${\boldsymbol \Gamma}_F+{\boldsymbol \Gamma}_{\fbar}={\boldsymbol \Gamma}$.
Similar to Eq.~\eqref{eq:det(S)}, Eq.~\eqref{eq:det(R)} relates the determinant of an $N_{\rm in}$-by-$N_{\rm in}$ matrix to a ratio of functional determinants.
In particular, Eq.~\eqref{eq:det(R)} shows that $\det{\bf R}_{\rm in}(\omega)$ is a rational function of $\omega$, 
which provides a rigorous basis for the topological-charge interpretation given earlier in Section~\ref{sec:R-zeros}.

Eq.~\eqref{eq:det(R)} is a central result
of this work.
It relates the R-zeros to a new effective operator
\begin{equation}\label{eq:A_RZ}
    \hat A_{\rm RZ}(\omega) \equiv \hat A_0(\omega) - \hat \Delta(\omega) - i[\hat \Gamma_F(\omega)-\hat \Gamma_ {\fbar}(\omega)].
\end{equation}
In particular, $\det {\bf R}_{\rm in}(\omega_{\rm RZ})=0$ at an R-zero, so we necessarily have $\det \hat A_{\rm RZ}(\omega_{\rm RZ}) = 0$, which defines a nonlinear eigenvalue problem
\begin{gather}
\label{eq:det(A)=0}
    \hat A_{\rm RZ}(\omega_{\rm RZ}) \ket{\omega_{\rm RZ}} = 0.
\end{gather}
We expect that there exists a countably infinite set of R-zeros at complex frequencies $\{\omega_{\rm RZ}\}$ that satisfy Eq.~\eqref{eq:det(A)=0} for each choice of input channels $F$, similar to resonances, which satisfy $\hat{A}_{\rm eff}(\omega_p) \ket{\omega_p} = 0$.
Like Eqs.~\eqref{eq:RSM definition} or~\eqref{eq:det_Rin_zero}, Eq.~\eqref{eq:det(A)=0} is universal and applicable to any open system with any choice of reflectionless input channels. 
Note that the denominator of Eq.~\eqref{eq:det(R)} is the same as that of Eq.~\eqref{eq:det(S)} while the numerators differ, implying that the poles (resonances) of ${\bf R}_{\rm in}$ are the same as those of ${\bf S}$, while zeros are different.
Implementation details for solving Eq.~\eqref{eq:det(A)=0} in different geometries are given later in Section~\ref{sec:RSM_wave_operator} \& Appendix~\ref{ap:boundary-matching}, where it is shown that the construction given here is equivalent to specifying mode-matched boundary conditions.

A caveat of using Eq.~\eqref{eq:det(A)=0} as opposed to Eq.~\eqref{eq:RSM definition} is that while every R-zero corresponds to an eigenmode of $\hat A_{\rm RZ}$, some eigenmodes of $\hat A_{\rm RZ}$ may not be R-zeros.
This can be seen from Eq.~\eqref{eq:det(R)}: $\det {\bf R}_{\rm in}$ is not generally zero if both the numerator and the denominator on the right-hand side are zero, so that we have a simultaneous eigenmode of both $\hat A_{\rm RZ}$ and $\hat A_{\rm eff}$.
Such simultaneous eigenmodes are rare, but the aforementioned BIC~\cite{2016_Hsu_NRM} is an example, as pointed out in Ref.~\cite{2018_Bonnet-BenDhia_PRSA} in a more specialized context.
In addition, resonances that do not radiate into certain channels~\cite{2014_Ramezani_PRL_2,2016_Zhou_Optica, 2019_Yin_arXiv}---sometimes referred to as ``unidirectional BICs''---are also such simultaneous eigenmodes.
BICs and unidirectional BICs typically require additional parameter tuning beyond the basic RSM problem~\cite{2016_Hsu_NRM} since they have both zero incoming {\it and} zero outgoing waves in the dark channels.
Nonetheless, they become more common when symmetries are present in the system.

We note the important intuition provided by Eqs.~\eqref{eq:det(R)}--\eqref{eq:det(A)=0}: for the purpose of reasoning about the zeros of ${\bf R}_{\rm in}$, each input channel acts as an effective ``irradiative gain'' to the system (because it contributes a negative semidefinite term to $\hat A_{\rm RZ}$, which then increases the imaginary part of its eigenvalue $\omega_{\rm RZ}$), and each output channel as an effective ``radiative loss''.
The balance of these two terms will determine the proximity of an R-zero to the real-frequency axis, where it becomes an RSM.
Hence when defining an R-zero problem in which few input channels scatter to many output channels, we expect the R-zero frequency to appear in the lower half of the complex plane for a passive system (similar to resonances).
Therefore to realize a steady-state solution we can either add gain in the dielectric function of the operator $\hat A_0$ (gain-loss engineering) or modify the scattering structure to increase the coupling of the input channels (index tuning) so as to move the solution to a real frequency.
In the opposite case of many more input channels than output channels, we expect the R-zero frequency to appear in the upper half plane and require adding absorption or increased coupling to the output channels to reach the real axis. 
This argument is merely qualitative, since the operator in Eq.~\eqref{eq:A_RZ} is the sum of non-commuting terms, so that the imaginary part of the eigenvalue is not simply the sum of contributions from each term.
However this qualitative picture is confirmed by the approximate coupled-mode analysis of Section~\ref{sec:RSM_TCMT} and by exact numerical solutions of several examples (see Figs.~\ref{fig:fp},\ref{fig:octopus},\ref{fig:free_space}).

There is an important difference between the zeros of ${\bf S}$ and of ${\bf R}_{\rm in}$.
Flux conservation means that the totally absorbed steady-state CPA cannot be realized without introducing absorption.
On the other hand, flux conservation does not prevent the existence of R-zeros on the real axis (RSMs) since the incident flux can be redirected to the $\fbar$ output channels.
Mathematically, ${\bf S}$ is unitary for lossless systems at real frequencies, with unimodular eigenvalues, but ${\bf R}_{\rm in}$ is not, and may have a zero even for real frequency.
This is why RSMs can be achieved via pure index tuning, even though CPA and lasing cannot; we will give explicit examples later [Fig.~\ref{fig:fp}(e), Fig.~\ref{fig:octopus}, Fig.~\ref{fig:free_space}].

\subsection{Symmetry Properties of R-zeros and RSMs \label{sec:Symmetry Properties}}

We now analyze exact properties of R-zeros and RSMs under time-reversal and discrete spatial transformations, in the presence of symmetries with respect to these transformations.
An earlier work~\cite{2013_Mostafazadeh_PRA} studied the consequences of symmetry in 1D, and has some overlap with what follows, though the present treatment generalizes to any dimension, and applies to complex R-zeros, not only real RSMs.

\subsubsection{Time-reversal Transformation (${\cal T}$) and Symmetry\label{sec:sym_T}}

The action of time reversal (${\cal T}$) is to complex conjugate everything, including the wave operators $\hat A$ and $\hat A_{\rm RZ}$, and the field ${\bf H}({\bf r})$. This interchanges the input and output channels ($F\leftrightarrow \fbar$):
\begin{equation}
    \mathcal{T}: (\omega,\varepsilon({\bf r}),F) \rightarrow (\omega^*,\varepsilon^*({\bf r}),\fbar).
    \label{eq:T-symmetry}
\end{equation}
It follows that if a cavity with dielectric function $\varepsilon ({\bf r})$ has an R-zero at frequency $\omega_0$ with input channels given by $F$, then the cavity with $\varepsilon^*({\bf r})$ has an R-zero at frequency $\omega_0^*$ with input channels given by the complement, $\fbar$.

\begin{figure}
    \centering
    \includegraphics[width=0.95\columnwidth]{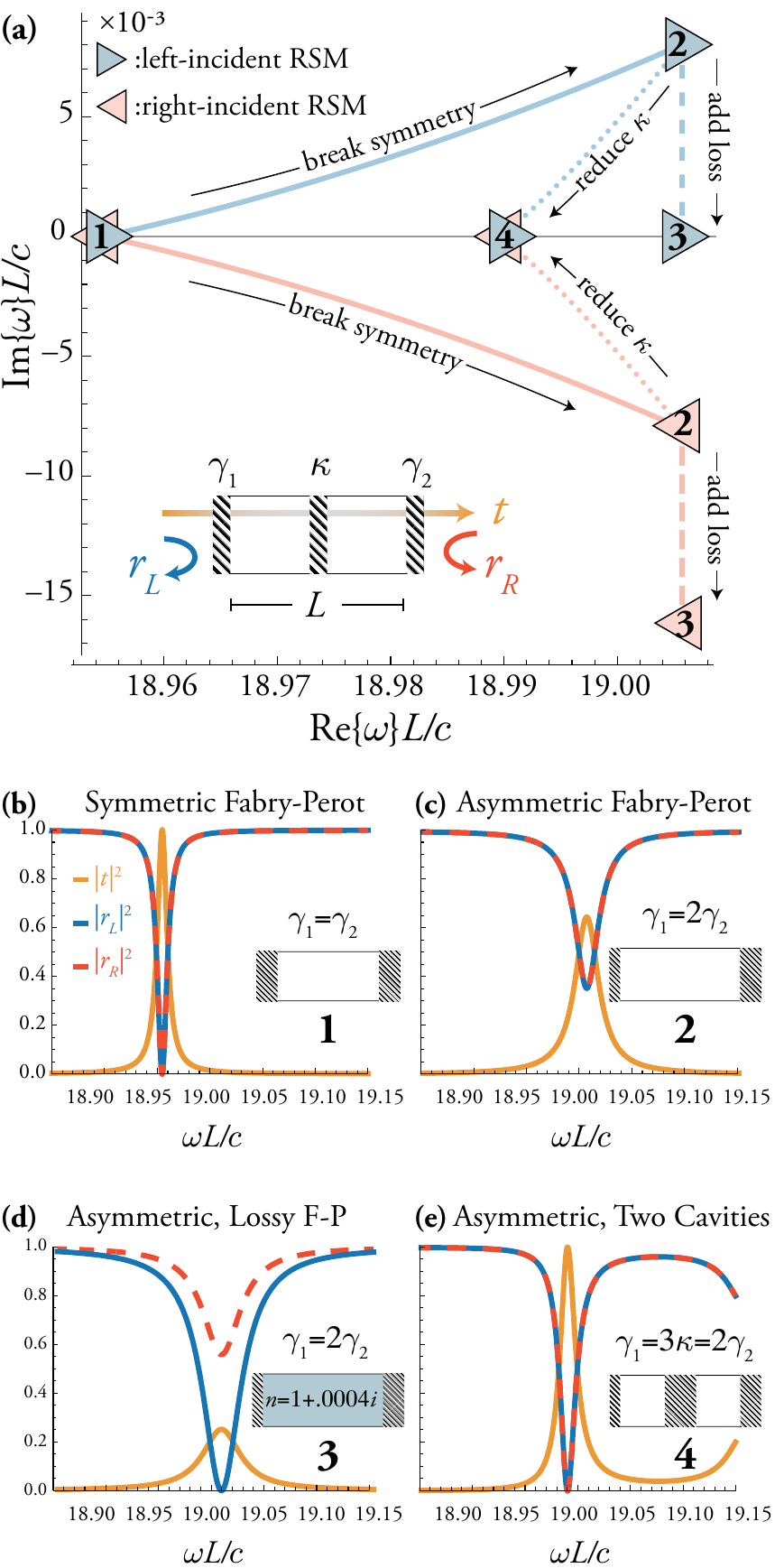}
    \caption{
    (Color) Illustration of RSMs and R-zero spectrum for simple two- and three-mirror resonators of length $L$ in 1D, consisting of $\delta$-function mirrors of strengths $\gamma_1^{-1}, \gamma_2^{-1}$ and $\kappa^{-1}$, as indicated in the schematic in (a).
    Throughout, we fix $\gamma_2 \equiv c/L$.
    Blue and red lines $1 \to 2$ indicate the effect of breaking symmetry by varying $\gamma_1$ from $\gamma_2 \to 2 \gamma_2$.
    A bidirectional RSM [as in (b)] splits into two complex-conjugate R-zeros off the real axis and a reflectionless steady-state (RSM) no longer exists, as in (c).
    Adding absorption to the cavity, indicated by blue and red lines $2 \to 3$, brings the upper R-zero to the real axis (but not the lower one), creating a unipolar left-incident lossy RSM, as in (d).
    Alternatively, adding a middle mirror and reducing its $\kappa$ from $\infty \to 2 \gamma_2/3$ is sufficient to bring both R-zeros back to the real axis ($2 \to 4$ in (a)), creating a bipolar RSM at a different frequency from the symmetric Fabry-P\'erot resonator (see (e)), without restoring parity symmetry.
    }
    \label{fig:fp}
\end{figure}

The cavity/scatterer is said to have time-reversal symmetry when $\varepsilon({\bf r})=\varepsilon^*({\bf r})$, i.e.,~when there is no absorption or gain. In such case, the ``two cavities'' described above are the same, and we can conclude that
\begin{itemize}
    \item Complementary R-zeros come in complex-conjugate pairs for flux-conserving (passive) cavities.
\end{itemize}
This is true independent of the presence or absence of spatial symmetries, such as parity.
The system can then be tuned to have an RSM either by index tuning or by gain-loss tuning. 
In the former case, the new cavity still exhibits time-reversal symmetry, so the resulting RSM is {\it bipolar}, by which we mean that a complementary RSM exists at the same frequency;
examples are given in Fig.~\ref{fig:fp}(b,e), Fig.~\ref{fig:octopus}, Fig.~\ref{fig:free_space}.
In the latter case, the new cavity no longer exhibits time-reversal symmetry, and we do not expect another RSM at that frequency; we refer to this as {\it unipolar}.

In Fig.~\ref{fig:fp}(a) we illustrate the concept of tuning to create an RSM for the simple case of an asymmetric Fabry-P\'erot cavity.
We start with a symmetric Fabry-P\'erot cavity, which has both $\mathcal{P}$ and $\mathcal{T}$ symmetry (this symmetry class will be discussed in more detail in Section~\ref{sec:sym_P_and_T}); it is well-known that such a cavity has equally-spaced unit-transmission resonances~\cite{1986_Siegman_book}, which we refer to as bidirectional RSMs (see Fig.~\ref{fig:fp}(b)).
We break parity symmetry but maintain the $\mathcal{T}$ symmetry by making the mirror reflectivities unequal (Fig.~\ref{fig:fp}(a) solid lines); as a consequence there is no longer an RSM for either direction of incidence at any real frequency (Fig.~\ref{fig:fp}(c)).
However, there is now a pair of complex-conjugate R-zeros off the real axis, as demanded by the $\mathcal{T}$ symmetry.
Here we have made the left barrier less reflective than the right one, and we observe that the left-incident R-zero has moved to the upper half plane and the right-incident R-zero to the lower half plane, which is consistent with the intuitions given in Section~\ref{sec:rigorous RSM}.  

One way to recover the RSMs without restoring parity symmetry is to use gain-loss engineering (non-hermitian tuning).
The intuitions of Section~\ref{sec:rigorous RSM} suggest that adding absorption will reduce the imaginary parts of the R-zero frequencies, which can be used to create a left-incident RSM.
A physical argument supporting this conclusion is that when the mirror reflectivities are equal, the prompt reflection from either mirror is cancelled by the total left/right reflection of waves reaching the interior and internally reflecting an odd number of times before escaping back in the incident direction.
When the left mirror is less reflective than the right one, its prompt reflection is decreased and the internal reflection backwards is increased, so total destructive interference is not possible.
However if one adds the right amount of absorption to the interior, these two amplitudes can again be balanced and destructive interference in the backwards direction can be restored.  This change has the opposite effect on a right input wave, making destructive interference impossible, and leading to a unipolar left-incident RSM.
For a right-incident RSM one would instead have to add the same amount of gain to the interior.
This physical picture is confirmed by Fig.~\ref{fig:fp}(a) dashed lines and Fig.~\ref{fig:fp}(d). 

Alternatively, we can employ index tuning to create an RSM, in this case by adding a third lossless mirror in the interior.
We can think of the left region and the interior mirror as forming a composite mirror such that for some mirror reflectivity and input frequency, left and right escape is again balanced.
Indeed such a three-mirror system does have an RSM with one-parameter tuning as shown by Fig.~\ref{fig:fp}(a) dotted lines and Fig.~\ref{fig:fp}(e).
Moreover, as the system is flux-conserving and reciprocal, it must be a bipolar RSM.
Comparison of Figs.~\ref{fig:fp}(a)(d)(e) illustrates this important difference between the two types of tuning: for the lossless case, both left- and right-incident R-zeros maintain their complex conjugate relation as the interior mirror is tuned and hence meet on the real axis.

    \subsubsection{Parity Transformation (${\cal P}$) and Symmetry\label{sec:sym_P}}

Consider a parity transformation ${\cal P}$ satisfying $\hat{\cal P}^2 = \hat{1}$ and $\det {\cal P} = -1$.
Common examples of parity are reflections (e.g.~$x \to -x$) and inversion in 3D ($x \to -x$, $y \to -y$, $z \to -z$).

Generally, the action of ${\cal P}$ is to leave the frequency unchanged and to map $\varepsilon({\bf r}) \rightarrow \varepsilon({\cal P}{\bf r})$, ${\bf H}({\bf r}) \rightarrow {\bf H}({\cal P}{\bf r})$, relating an R-zero/RSM of one structure to that of a structure transformed by ${\cal P}$.
However, these two R-zeros are generally not complementary to each other.
Therefore, we further require the input-output channels to have ${\cal P}$ symmetry, by which we mean that $\varepsilon({\bf r}) = \varepsilon({\cal P}{\bf r})$ in the asymptotic region and that the input channels are chosen such that $F$ maps to $\fbar$ under ${\cal P}$; we call this a ``bisected'' partition of the channels.
The most common bisected systems would be those that naturally divide into ``left'' and ``right'' and for which the input channels are chosen to be all left or all right channels,
which generalizes the well-studied one-dimensional case.
But there are also other possibilities, e.g.~a partition into clockwise and couterclockwise channels for a finite-sized scatterer in free space.
For a bisected partition $F$, the action of ${\cal P}$ is
\begin{equation}
    \mathcal{P}: (\omega,\varepsilon({\bf r}),F) \rightarrow (\omega,\varepsilon({\cal P}{\bf r}),\fbar).
\end{equation}
When an R-zero with a bisected partition is bipolar, we refer to it as {\it bidirectional}; when it is unipolar, we refer to it as {\it unidirectional}.

We say that the cavity/scatterer has ${\cal P}$ symmetry when $\varepsilon({\bf r}) = \varepsilon({\cal P}{\bf r})$, for which we can say that
\begin{itemize}
    \item When the cavity and the channel partition both have $\mathcal{P}$ symmetry, R-zeros are bidirectional, whether or not $\omega_{\rm RZ} \in\mathbb{R}$.
\end{itemize}
This is in contrast to the case with $\mathcal{T}$-symmetry, where bipolarity only holds for real-frequency RSMs.

    \subsubsection{Parity-Time Transformation (${\cal PT}$), Symmetry, and Symmetry-Breaking Transition\label{sec:sym_PT}}

The action of the joint parity-time ($\mathcal{PT}$) operator, i.e performing both $\mathcal{P}$ and $\mathcal{T}$ transformations simultaneously, is particularly interesting as it is the case of $\mathcal{PT}$ symmetry that brought recent attention to unidirectional RSMs.
We assume that the asymptotic region has $\mathcal{P}$ symmetry and that the partition $F$ is bisected; hence the action of $\mathcal{PT}$ is
\begin{equation}
    \mathcal{PT}: (\omega,\varepsilon({\bf r}),F) \rightarrow (\omega^*,\varepsilon^*({\cal P}{\bf r}),F).
\end{equation}
When the scattering region has $\mathcal{PT}$-symmetry, namely when
$\varepsilon^*({\cal P}{\bf r}) = \varepsilon({\bf r})$, we have
\begin{itemize}
    \item Systems with ${\cal PT}$-symmetry either have unidirectional RSMs with $\omega_0\in\mathbb{R}$, or complex-conjugate pairs of unidirectional R-zeros of the {\it same directionality} with $\omega_0\in\mathbb{C}$.
\end{itemize}
This is in contrast to the case of ${\cal T}$-symmetry alone, which allows real bidirectional RSMs or complex-conjugate pairs of R-zeros of {\it opposite} polarity/directionality.

\begin{figure}
   \centering
   \includegraphics[width=0.875\columnwidth]{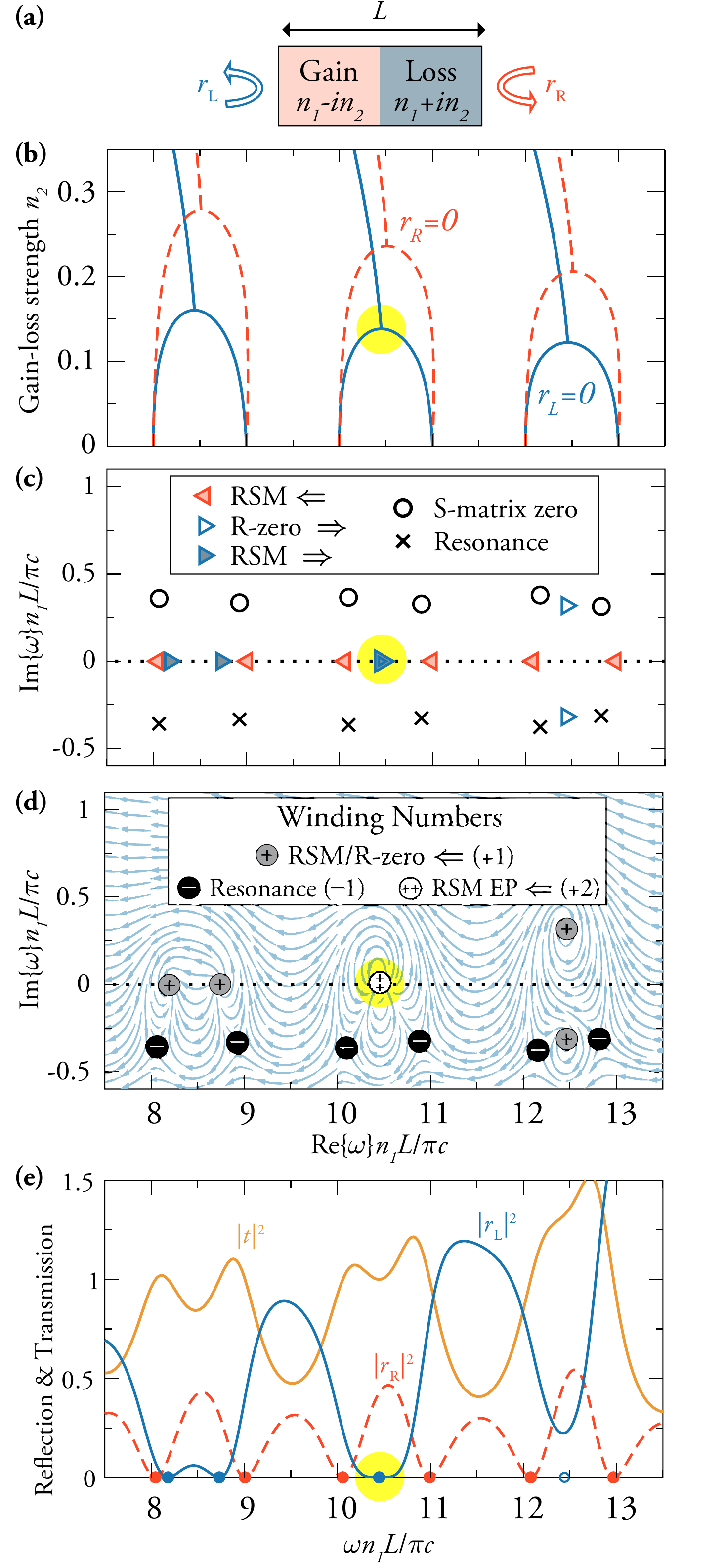}
   \caption{
   (Color) RSMs in a $\mathcal{PT}$-symmetric structure.
   {\bf (a)} Schematic of the structure: an etalon in air, with refractive index $n = n_1 - i n_2$ on the left and $n = n_1 + i n_2$ on the right; here $n_1=2$.
   {\bf (b)} Real part of the R-zero frequencies (blue solid lines: left-incident; red dashed lines: right-incident) as the gain-loss strength is increased.
   For small $n_2$, the frequencies are real valued.
   After two RSMs meet at an RSM EP (highlighted in yellow), they split into two R-zeros at complex-conjugate frequencies.
   {\bf (c)} Spectra of the R-zeros/RSMs, S-matrix zeros, and resonances in the complex-frequency plane at $n_2 = 0.13844$ where two right-going RSMs meet.
   {\bf (d)} Streamlines of the vector field ${\boldsymbol(}{\rm Re}(r_{\rm L}), {\rm Im}(r_{\rm L}){\boldsymbol)}$, showing $+1$ topological charges at R-zeros/RSMs and $-1$ topological charges at resonances. When two RSMs meet at an RSM EP, the topological charges add up to $+2$.
   {\bf (e)} Reflection and transmission spectra at the same $n_2$ as in (c,d); blue and red filled dots mark the RSM frequencies, open blue dot is real part of complex R-zero, which has already crossed the threshold.
   }
\label{fig:PT}
\end{figure}

This property of the ${\cal PT}$ case implies something quite important.
Often ${\cal PT}$ scattering systems are studied by beginning with a flux-conserving system (satisfying both ${\cal P,T}$ symmetries separately) and adding gain and absorption anti-symmetrically so as to break ${\cal P}$ and ${\cal T}$ symmetries while preserving $\mathcal{PT}$.
The initial system has bidirectional RSMs, i.e.,~a degenerate pair of left and right RSMs (see discussion of the ${\cal P,T}$ case below in Section~\ref{sec:sym_P_and_T}) but has no other degeneracy; thus the left and right RSMs are constrained to remain on the real axis as {\it unidirectional} RSMs when the value of the gain-loss strength increases.
(If, e.g.,~a left-incident RSM moved off the real axis immediately, it would lack a second left incident RSM as complex-conjugate partner, violating the condition above.)
These unidirectional RSMs are invariant under the ${\cal PT}$ transformation, so they are said to be in the ${\cal PT}$-unbroken phase.
As the gain-loss strength is further increased, eventually each RSM may meet another RSM of the same directionality at a real frequency (see Fig.~\ref{fig:PT}(b) below as an example), above which point, generically, the pair of RSMs will leave the real axis as complex-conjugate pairs of R-zeros, becoming inaccessible in steady-state.
These complex-conjugate pairs of R-zeros do not exhibit ${\cal PT}$ symmetry (one maps to its conjugate partner under the ${\cal PT}$ transform), so they are said to be in the ${\cal PT}$-broken phase.
The above explains the commonly observed RSMs in 1D systems with ${\cal PT}$ symmetry~\cite{2011_Lin_PRL,2011_Hernandez-Coronado_PLA, 2011_Longhi_JPA, 2012_Ge_PRA, 2012_Regensburger_Nature, 2013_Feng_nmat, 2014_Ramezani_PRL_2} or anti-$\mathcal{PT}$ symmetry~\cite{2013_Ge_PRA} and their disappearance at large gain-loss parameters~\cite{2011_Hernandez-Coronado_PLA,2011_Longhi_JPA,2012_Ge_PRA,2012_Jones_JPA,2013_Mostafazadeh_PRA}.
It should be noted that the critical parameters that define the ${\cal PT}$ transitions are different for each RSM pair and different for each directionality.
They are also different from the transitions where a pair of unimodular eigenvalues of the S-matrix turn into an amplifying one and an attenuating one~\cite{2011_Chong_PRL_1, 2012_Ge_PRA}.

This behavior is illustrated for the one-dimensional ${\cal PT}$-symmetric structure shown in Fig.~\ref{fig:PT}(a): an etalon of thickness $L$ in air where the left half has refractive index $n = n_1 - i n_2$ and the right half has index $n = n_1 + i n_2$.
For a passive etalon ($n_2=0$), bidirectional RSMs exist at real  frequencies $\omega_{\rm RSM} = m \pi c/n_1L$ with $m\in{\mathbb Z}$.
When the gain-loss strength $n_2$ is increased, pairs of RSMs move toward each other in frequency as shown in Fig.~\ref{fig:PT}(b).
As parity and time-reversal symmetries are individually broken, the right-incident RSMs (for which $r_{\rm R}=0$) and the left-incident RSMs (for which $r_{\rm L}=0$) now occur at different frequencies.
But since the system still exhibits $\mathcal{PT}$ symmetry, all of these RSM frequencies remain real-valued.
At critical values of $n_2$, a pair of RSMs meet. 
As $n_2$ is further increased, the pair of RSMs split into two, leaving the real-frequency axis as complex-conjugate pairs of R-zeros. 
The RSM spectrum in the complex-frequency plane, together with the S-matrix zeros and poles (resonances), is shown in Fig.~\ref{fig:PT}(c) for a critical value of $n_2$ where two right-going RSMs meet.
The corresponding reflection and transmission spectra are given in Fig.~\ref{fig:PT}(e).

We use this example to illustrate the topological properties of R-zeros/RSMs and resonances.
Fig.~\ref{fig:PT}(d) shows the streamlines of the vector field ${\boldsymbol (}{\rm Re}(r_{\rm L}), {\rm Im}(r_{\rm L}){\boldsymbol )}$ in the complex-frequency plane.
Following the discussion in Section~\ref{sec:R-zeros}, we indeed observe that $\arg(r_{\rm L})$ winds by $2\pi$ along counterclockwise loops around each R-zero or RSM (corresponding to a $+1$ topological charge), while it winds by $-2\pi$ around each resonance (corresponding to a $-1$ topological charge).
When two RSMs meet, they superpose as one topological defect with charge $+2$, as highlighted in yellow.
We further discuss such a coalescence of RSMs in the next section.

    \subsubsection{RSM Exceptional Points\label{sec:RSM EP}}

Non-hermitian operators have the property that when two eigenvalues become degenerate (in both their real and imaginary parts), the two associated eigenvectors also coalesce into one.
Such a coalescence is called an exceptional point (EP) in parameter space and has many unique properties~\cite{1995_Kato_book,2012_Heiss_JPA,2017_Feng_nphoton,2018_ElGanainy_nphys,2019_Miri_Science,2019_Ozdemir_nmat}.
EPs of the purely outgoing (resonance) wave operator have been widely studied~\cite{2000_Persson_PRL,2009_Rotter_JPA,2012_Liertzer_PRL,2014_Brandstetter_NatComm,2014_Wiersig_PRL,2015_Zhen_Nature,2016_Peng_PNAS,2017_Chen_Nat,2017_Hodaei_Nat,2018_Zhou_Science,2018_Zhang_Arxiv}, and recently so too have EPs of the purely incoming wave operator with ${\cal PT}$-symmetry~\cite{2017_Achilleos_PRB} or unconstrained by any symmetry~\cite{2019_Sweeney_PRL}.

The R-zero/RSM wave operator $\hat A_{\rm RZ}$ of Eq.~\eqref{eq:A_RZ} is non-hermitian and shares many similarities with the purely outgoing (resonance) wave operator $\hat A_{\rm eff}$.
It is therefore possible to create {\it RSM exceptional points} where multiple reflectionless states coalesce into one, which
is, from a physical point of view, a new kind of EP not previously studied (to our knowledge).  
The aforementioned RSM transitions in ${\cal PT}$ systems are examples of this.
These RSM EPs share some common features with the perfectly absorbing EPs of Refs.~\cite{2017_Achilleos_PRB,2019_Sweeney_PRL}.
In particular, there is no self-oscillating instability (lasing) when the EP reaches the real axis; a steady-state RSM EP is compatible with linear response.
At an RSM EP, the lineshape of the reflection intensity will change from its generic quadratic form to a quartic, flat-bottomed lineshape, characteristic of a $+2$ topological charge.
An example of this effect is shown in Fig.~\ref{fig:PT}(d--e), where the RSM EP is highlighted in yellow.
Near an EP of any kind, the complex eigenvalue typically exhibits a square-root dependence on system parameters, visible in Fig.~\ref{fig:PT}(b).

Note that RSMs in one-dimensional systems have been characterized in some works as EPs of an unconventional non-symmetric scattering matrix~\cite{2007_Cannata_AoP,2013_Feng_nmat,2013_Mostafazadeh_PRA,2014_Wu_PRL,2014_Zhu_OE,2014_Kang_PRA,2017_Huang_nanoph}.
However, those are non-degenerate RSMs and are not EPs of the underlying wave operator.
Also, in higher dimensions, an EP of this unconventional non-symmetric scattering matrix is no longer a reflectionless state.
Therefore, we do not adopt this convention and reserve the term ``RSM EP'' for the states discussed in this section.

    \subsubsection{${\cal P}$ and ${\cal T}$ Symmetry and Symmetry-breaking Transition\label{sec:sym_P_and_T}}

The final symmetry class we will discuss is the case of systems with both ${\cal P}$ and ${\cal T}$ symmetries.
The symmetric Fabry-P\'erot cavity, discussed briefly above, is a familiar example.
Such a ${\cal P,T}$ system simultaneously exhibits all of the symmetry properties given in Sec.~\ref{sec:sym_T}--\ref{sec:sym_PT}.
Therefore we can expect bidirectional RSMs on the real-frequency axis without any tuning, with flux conservation implying unit transmission for these RSMs.
The only other possibility allowed by symmetry is bidirectional complex-conjugate pairs of R-zeros, which is less familiar.

We have confirmed that this possibility does occur in a physical model by starting with a ${\cal P,T}$ resonator with bidirectional unit-transmission RSMs, and by tuning parameters of the cavity (index tuning) while preserving both symmetries, we find that we are able to induce a symmetry-breaking transition in which pairs of bidirectional RSMs meet on the real axis at two degenerate EPs, and then move off the real axis as bidirectional complex-conjugate pairs of R-zeros. 
Such degenerate RSM EPs are examples of steady-state exceptional points in a flux-conserving system, which has never been demonstrated before to our knowledge. Moreover, one can show that this transition is actually associated with ${\cal PT}$-symmetry breaking, despite the absence of gain or absorption.
We will defer detailed discussion of this interesting case to a future article~\cite{2019_Sweeney}.

\begin{figure}
    \centering
    \includegraphics[width=0.95\columnwidth]{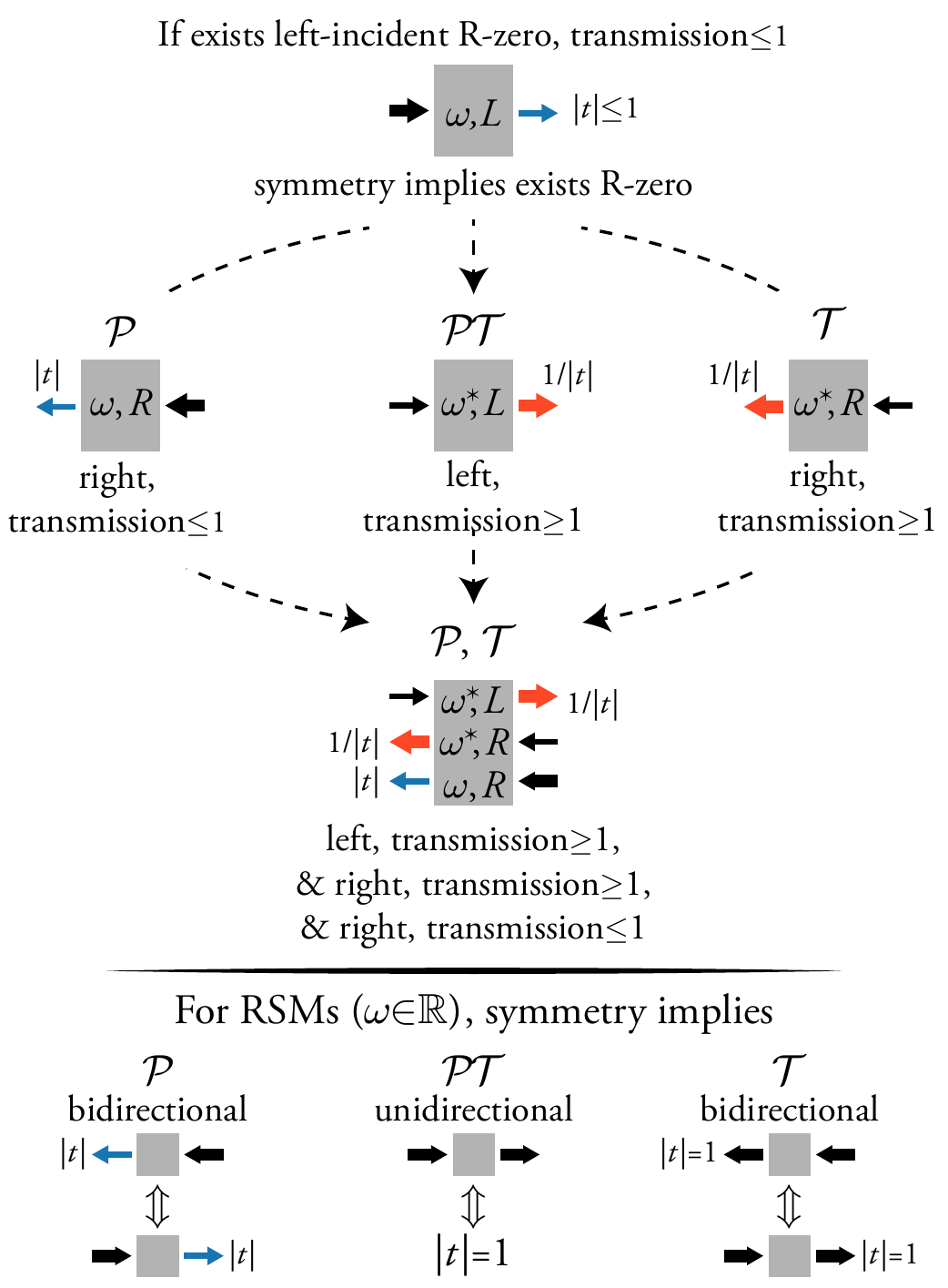}
    \caption{
    (Color) Schematic illustrating the implications of symmetry for R-zeros and RSMs.
    (Top) Beginning with the assumption of the existence of a left-incident R-zero with with transmission $|t|<1$ at some frequency, $\omega$, there will exist other R-zeros with specific properties based on the presence of $\mathcal{P,T,PT}$ symmetry (as shown by arrows).
    (Bottom) The initial state has real frequency (is an RSM), which exists in steady-state; the implications of the various symmetries are shown for this case.
    Note that if $\omega$ is real, the implications for the $\mathcal{T}$ and ($\mathcal{P,T}$) are the same: bidirectional unit transmission reflectionless states.
    }
    \label{fig:symmetry}
\end{figure}

The different symmetry classes and their properties are summarized in Table~\ref{tab:Symmetry Table} and Fig.~\ref{fig:symmetry}.

\begin{table*}[]
    \centering
    \begin{tabular}{c||c|c|c|c|c|c}
     \thead{symmetry\\type} & $\eps{x}=\ $ & $\omega_F\Rightarrow\phantom{\omega}$ & RSM ($\omega_F\in{\mathbb R}$) & $N_{\rm RSM}$ & $N_{\rm RSM}^{\rm EP}$ &  \thead{$\mathbb{R}\rightarrow \mathbb{C}$\\transition?} \\ \hline\hline
     none & any &  & unipolar & 1 & 3 & \\ \hline
     ${\mathcal T}$ & $\varepsilon^*({\bf x})$ & $\omega_{\fbar}^*$ & bipolar & 1 & 3 & \\ \hline
     ${\mathcal P}$ & $\eps{{\cal P} x}$ & $\omega_{\fbar}$ &  bidirectional & 1 & 3 & \\ \hline
     ${\cal PT}$ & $\varepsilon^*({\cal P}{\bf x})$ & $\omega_F^*$ & unidirectional & 0/1 & 1 &  \checkmark \\ \hline
     ${\cal P,T}$ & $\varepsilon^*({\bf x}),\ \eps{{\cal P}x}$ & $\omega_{\fbar},\ \omega^*_{\fbar},\ \omega^*_F$ & bidirectional & 0/1 & 1 & \checkmark
    \end{tabular}
    \caption{
    Consequences of discrete symmetries for the RSM problem.
    $F$ is the set of input states.
    Unipolar means that only one set of input channels has zero reflection at a given frequency and bipolar means that its complement does too.
    In this table we assume that the channels are bisected, i.e., the set of input channels maps to its complement under a parity transformation ${\cal P}$.
    In this case we use the terms unidirectional instead of unipolar, bidirectional instead of bipolar.
    $N_{\rm RSM}$ is the minimum number of system parameters that must be tuned (consistent with symmetry) to achieve an RSM (i.e., make $\omega_F$ real), while $N_{\rm RSM}^{\rm EP}$ parameters are necessary for an exceptional point of RSMs (i.e. degenerate, real $\omega_{\rm RZ}$).
    }
    \label{tab:Symmetry Table}
\end{table*}

    \subsection{Coupled-mode Analysis\label{sec:RSM_TCMT}}

The preceding results were derived directly from the Maxwell's equations and involve no approximation.
Meanwhile, in many circumstances an approximate analtyic model will be adequate and desirable for its simplicity.
In photonics a standard tool is the temporal coupled-mode theory (TCMT)~\cite{Haus_book, 2003_Fan_JOSAA, 2004_Suh_JQE, 2018_Wang_OL, 2019_Zhao_PRA, 2017_Alpeggiani_PRX}, which is a phenomenological model widely used in the design and analysis of optical devices~\cite{2012_Verslegers_PRL, 2014_Peng_nphys, 2014_Hsu_NL, 2015_Zhen_Nature}.
The TCMT formalism is derived from symmetry constraints~\cite{Haus_book, 2003_Fan_JOSAA, 2004_Suh_JQE, 2018_Wang_OL, 2019_Zhao_PRA} rather than from first principles, yet it leads to an analytic relation between the scattering matrix and the underlying Hamiltonian that is similar to Eq.~\eqref{eq: Heidelberg} and is reasonably accurate in many cases.
Assuming harmonic time dependence, a reciprocal cavity or scattering region supporting $M$ internal resonances and coupled to $2N$ external channels ($N$ incoming and $N$ outgoing) can be described by~\cite{2004_Suh_JQE}
\begin{subequations}
\label{eq:TCMT}
\begin{align}
    \label{eq:TCMT_1}
    -i \omega {\bf a}  &=  -i {\bf H}_{\rm eff} {\bf a}  + {\bf D}^T {\boldsymbol \alpha} , \\
    \label{eq:TCMT_2}
    {\boldsymbol \beta} &= {\bf S}_0 {\boldsymbol \alpha} + {\bf D} {\bf a} , 
\end{align}
\end{subequations}
where ${\bf H}_{\rm eff}$ is an $M$-by-$M$ effective Hamiltonian
\begin{equation}
    \label{eq:H_eff}
    {\bf H}_{\rm eff} \equiv {\bf H}_{\rm close} - i \frac{{\bf D}^\dagger {\bf D}}{2},
\end{equation}
and ${\bf a} $ is a column vector containing the field amplitudes of the $M$ resonances.
The $N$-by-$M$ matrix ${\bf D}$ contains the coupling coefficients between the resonances and the channels; the $m$-th column of ${\bf D}$ is essentially the radiation wavefront of the $m$-th resonance analyzed in the channel basis.
The Hamiltonian matrix ${\bf H}_{\rm close}$ describes a closed system and is Hermitian in the absence of absorption or gain; $(\omega-{\bf H}_{\rm close})$ is analogous to ${\bf A}_0$ in Eq.~\eqref{eq:A0} or ${\bf  A}_0^\prime-{\boldsymbol \Delta}$ in Eq.~\eqref{eq:Aeff_and_Sigma}; the frequency shifts ${\boldsymbol \Delta}$ due to openness are not separately included in TCMT.
The positive semi-definite matrix ${\bf D}^\dagger {\bf D} / 2$ is analogous to ${\bf \Gamma}$ in Eq.~\eqref{eq:Aeff_and_Sigma}; its diagonal elements are the decay rates of the modes due to radiation into channels in the open environment, and its off-diagonal elements are the dissipative via-the-continuum coupling rates.
The ``direct scattering matrix'' ${\bf S}_0 = {\bf S}_0^T$ is symmetric and describes the ``non-resonant'' part of the scattering process that varies slowly with frequency.
The distinction between ``resonant'' and ``non-resonant'' is not always clear; the typical practice in TCMT is to use ${\bf H}_{\rm close}$ and ${\bf D}$ to model one or a few high-quality-factor resonances in the frequency range of interest, and bundle the contributions from the further-away and/or low-quality-factor resonances into ${\bf S}_0$ in an empirical manner.
In a passive system without absorption or gain, time-reversal symmetry requires that ${\bf S}_0 {\bf D}^* = -{\bf D}$ (Ref.~\cite{2004_Suh_JQE, 2019_Zhao_PRA}); here we assume that the possible presence of absorption or gain can be modeled through an anti-Hermitian term in ${\bf H}_{\rm close}$ without breaking the ${\bf S}_0 {\bf D}^* = -{\bf D}$ condition.
It follows that the scattering matrix is
\begin{equation}
\label{eq:S_form1}
{\bf S}(\omega) = \left( {\bf I}_N - i{\bf D} \frac{1}{\omega - {\bf H}_{\rm eff}} {\bf D}^{\dagger} \right) {\bf S}_0,
\end{equation}
where ${\bf I}_N$ is the $N$-by-$N$ identity matrix.
When all contributing resonances (including the low-quality-factor ones that may be far away in frequency) are included in ${\bf H}_{\rm close}$ and ${\bf D}$, the direct scattering matrix will be simply ${\bf S}_0= {\bf I}_{N}$ (Ref.~\cite{2017_Alpeggiani_PRX}).
The similarity between Eq.~\eqref{eq: Heidelberg} and Eq.~\eqref{eq:S_form1} is evident.
A formalism mathematically equivalent to TCMT is also used in quantum noise theory~\cite{2004_Gardiner_book}.

From Eq.~\eqref{eq:S_form1}, one may proceed to derive an expression for the determinant of the generalized reflection matrix similar to Eq.~\eqref{eq:det(R)}; we leave such an exercise to the interested readers, and instead provide an alternative approach here.
Let us define ${\bf D}_{\rm in} \equiv {\bf F} {\bf D}$, using the filtering matrix ${\bf F}$ introduced earlier, as the coupling coefficients into the $N_{\rm in}$ input channels defining $F$, and similarly ${\bf D}_{\rm out}$ as the coefficients to the output channels.
Consider ${\bf S}_0= {\bf I}_{N}$.
Using Eq.~\eqref{eq:Rin_S}, Eq.~\eqref{eq:S_form1}, and the Woodbury matrix identity~\cite{1950_Woodbury_MR}, we can write the inverse of the generalized reflection matrix as
\begin{equation}
\label{eq:S_in_inverse}
{\bf R}_{\rm in}^{-1}(\omega)
 = {\bf I}_{N_{\rm in}} + i {\bf D}_{\rm in} \frac{1}{\omega-{\bf H}_{\rm RZ}} {\bf D}_{\rm in}^{\dagger},
\end{equation}
where we have defined a matrix
\begin{equation}
\label{eq:H_RSM}
{\bf H}_{\rm RZ}
\equiv {\bf H}_{\rm close} + i \frac{{\bf D}_{\rm in}^\dagger {\bf D}_{\rm in}}{2} - i \frac{{\bf D}_{\rm out}^\dagger {\bf D}_{\rm out}}{2}.
\end{equation}
The matrix $(\omega-{\bf H}_{\rm RZ})$ is analogous to $\hat A_{\rm RZ}(\omega)$ in Eq.~\eqref{eq:A_RZ}.
At the frequency $\omega=\omega_{\rm RZ}$ of an R-zero, $ \det({\bf R}_{\rm in}^{-1}) = 1/\det({\bf R}_{\rm in})$ diverges, and Eq.~\eqref{eq:S_in_inverse} shows that such divergence can only happen when $\omega$ is an eigenvalue of ${\bf H}_{\rm RZ}$.
Therefore, every R-zero is necessarily an eigenmode of ${\bf H}_{\rm RZ}$.
Note, however, that the reverse is not true: not every eigenmode of ${\bf H}_{\rm RZ}$ is an R-zero, since it is possible for $|| {\bf D}_{\rm in}(\omega - {\bf H}_{\rm RZ})^{-1}{\bf D}_{\rm in}^\dagger||<\infty$ even when $\omega$ is an eigenvalue of ${\bf H}_{\rm RZ}$; this happens when the eigenmode ${\bf a}$ satisfies ${\bf D}_{\rm in} {\bf a} = 0$, which is precisely when it is a BIC or a one-sided resonance---see discussions at the end of Section~\ref{sec:rigorous RSM}.

Like $\hat A_{\rm RZ}$, the matrix ${\bf H}_{\rm RZ}$ can be understood intuitively.
Outgoing radiation into the $N_{\rm out}$ output channels introduces an effective radiative loss term $- i {\bf D}_{\rm out}^\dagger {\bf D}_{\rm out}/2$.
Incident light coming from the $N_{\rm in}$ input channels introduces an effective ``irradiation gain'' term $+i {\bf D}_{\rm in}^\dagger {\bf D}_{\rm in}/2$.

The case when there is only one dominant resonance in the frequency range of interest ($M=1$) is particularly instructive.
Here, ${\bf H}_{\rm close} = \omega_0 - i\gamma_{\rm nr}$ is a scalar where $\gamma_{\rm nr}$ is the non-radiative decay rate due to absorption or gain, and
the R-zero frequency given by Eq.~\eqref{eq:H_RSM} is
\begin{equation}
\begin{aligned}
\label{eq:omega_RSM_single_mode}
&\omega_{\rm RZ} = \left( \omega_0-i\gamma_{\rm nr} \right) + i \left( \gamma_{\rm in} - \gamma_{\rm out} \right), \\
&\gamma_{\rm in} \equiv \sum_{n \in F}|d_n|^2/2,
\quad \gamma_{\rm out} \equiv \sum_{n \notin F}|d_n|^2/2,
\end{aligned}
\end{equation}
where $d_n$ is the coupling coefficient (partial width) of the mode to the $n$-th channel.

In this single-mode approximation, which is widely used in the context of high-Q resonant structures, the intuitive understanding of RSMs is manifestly realized. 
The total in-coupling rate for the input channels acts as an effective source of gain, while the total decay rate into the output channels plus the intrinsic absorption in the cavity act as an effective loss.
When these two quantities are equal (critical coupling), the reflectionless state has a real frequency and becomes an RSM.
Within this approximation there are no multi-resonance pushing or pulling effects or other sources of real frequency shifts.
Therefore all the different R-zero boundary conditions simply move the frequency of the corresponding R-zero vertically in the complex plane between the purely outgoing solution (resonance) in the lower-half plane and the purely incoming solution (S-matrix zero) in the upper-half plane.
(The picture is trivially changed in the presence of absorption in the cavity, as the S-matrix resonance and zero move rigidly down and are no longer symmetric around the real axis.)
It is straightforward to also show that the RSM incident wavefront ${\boldsymbol \alpha}_{\rm RSM}$ is simply the phase conjugation of the resonance's radiation wavefront into the designated channels.
When $\omega = \omega_{\rm RZ}$, the non-resonant reflection ${\boldsymbol \alpha}_{\rm RSM}$ is exactly cancelled by the resonant scattering ${\bf D} {\bf a}$ back into those channels.
When the frequency is detuned from $\omega_{\rm RZ}$, the reflection signal rises as a Lorentzian function with the linewidth being that of the underlying resonance, $\left( \gamma_{\rm in} + \gamma_{\rm out} + \gamma_{\rm nr} \right)$.

The single-resonance scenario is the simplest example of an R-zero, yet it already explains the impedance-matching conditions previously found using TCMT in waveguide branches~\cite{2001_Fan_JOSAB}, antireflection surfaces~\cite{2014_Wang_Optica}, and polarization-converting surfaces~\cite{2017_Guo_PRL}: i.e.~that zero reflection in a passive ($\gamma_{\rm nr}=0$) system is achieved at $\omega_0$ when the decay rate into the incident channel $\gamma_{\rm in}$ equals the sum of decay rates into all outgoing channels $\gamma_{\rm out}$.

\begin{figure}
    \centering
    \includegraphics[width=0.95\columnwidth]{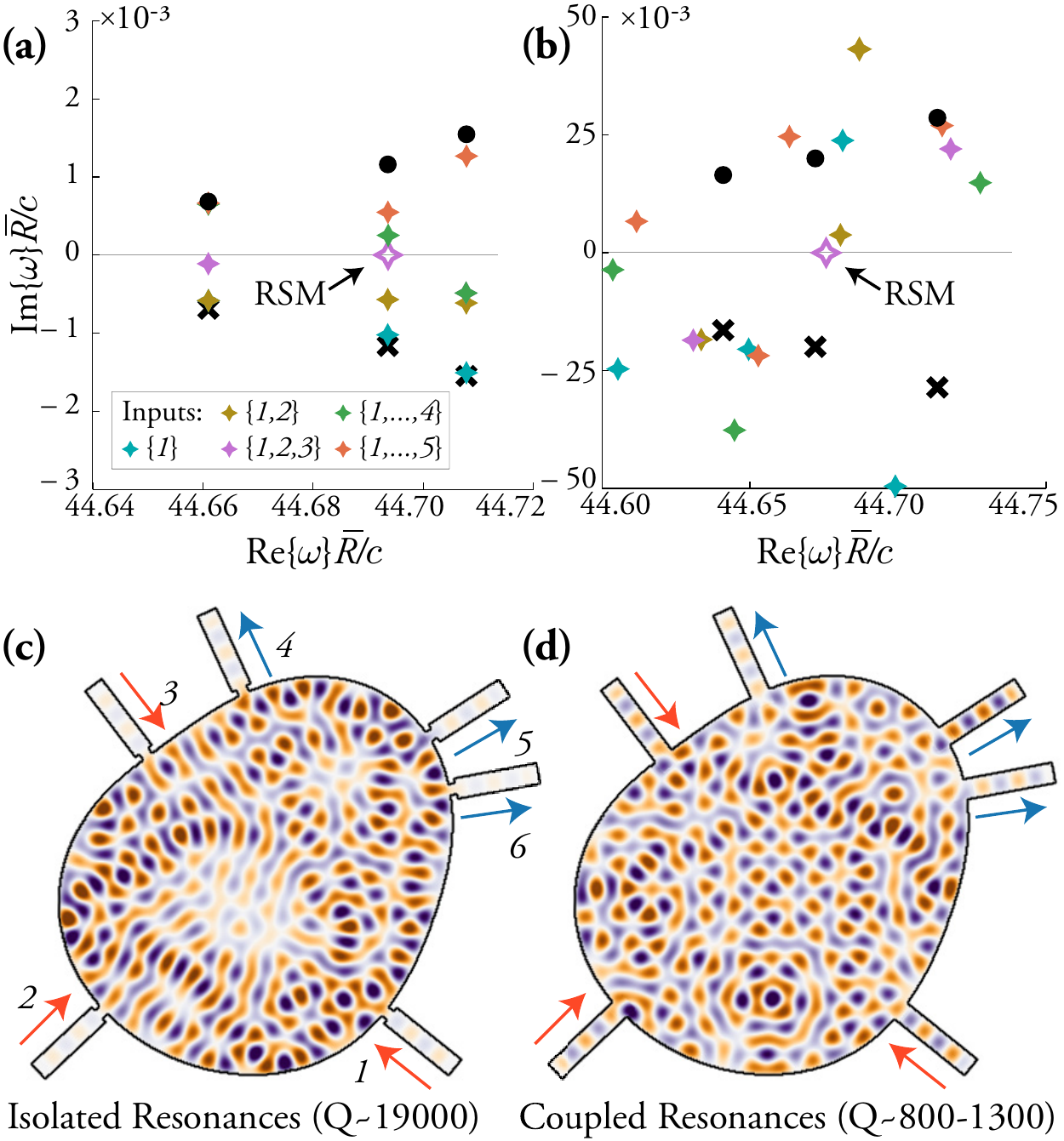}
    \caption{
    (Color) Asymmetric lossless waveguide junction/resonator (mean radius $\bar R$) coupled to six single-mode waveguides, with constrictions at the entrances to the junction.
    {\bf (a)} Numerically calculated R-zero spectrum for a weakly coupled, high-Q junction with well isolated resonances.
    Black x and dot are purely outgoing (resonance) and incoming (S-matrix zero) frequencies, which are complex conjugates.
    Colored stars are R-zeros for various choices of input channels; the legend indicates which channels are inputs, with the channel labels given in (c).
    The R-zeros cluster vertically above the resonance frequency and below the S-matrix zero frequency, as predicted by single resonance TCMT approximation Eq.~\eqref{eq:omega_RSM_single_mode}.
    The common width of the constrictions for waveguides $\{4,5,6\}$ is slightly tuned to make a 3-in/3-out R-zero real, creating an RSM.
    {\bf (b)} R-zero spectrum for the same junction but with the constrictions opened, which lowers the Q of the resonances (note change in vertical scale).
    The linewidths of the resonances are now comparable to their spacing. 
    Due to multi-resonance effects, the R-zeros are spread out along the real and complex frequency axis and are no longer associated with a single resonance.
    Nonetheless, by slightly tuning the constriction width as before, a 3-in/3-out R-zero is again made real (RSM), as in the high-Q case.
    {\bf (c--d)} The mode profiles of the RSMs for the high-Q (c) and low-Q (d) cases.
    }
    \label{fig:octopus}
\end{figure}

The single-resonance approximation is typically valid when a cavity has a high quality factor (Q) and is weakly-coupled to the input/output channels, so that its resonances are near the real axis and multi-resonance effects can be neglected at most frequencies.
However the general RSM theory applies equally well to low-Q cavities where we expect the simple picture just described to break down substantially. 
An example of both limits is given in Fig.~\ref{fig:octopus}, where we study, using an exact numerical method, an asymmetric resonant cavity connected to six single-mode waveguides.
The cavity shown in Fig.~\ref{fig:octopus}(c) has constrictions at its waveguide junctions to increase the quality factors of the resonances.
In Fig.~\ref{fig:octopus}(a) we show the R-zero spectrum for this cavity, which has the vertical clustering bracketed by the resonance and S-matrix zero, as predicted by the single resonance approximation just discussed.
In contrast, the cavity shown in Fig.~\ref{fig:octopus}(d) has the constrictions opened, which increases the quality factors by a factor of $\sim 20$.
For this case the single-resonance approximation fails, and the R-zero spectrum [Fig.~\ref{fig:octopus}(b)] is spread out in both the real and imaginary frequency axes.
Strikingly, some R-zeros lie below the resonances while others lie above the S-matrix zeros in the complex-frequency plane, something which is forbidden within the single resonance approximation.
Nonetheless, in both cases, we were able to tune to a real-frequency RSM for the three-in, three-out boundary condition simply by slightly varying the the constrictions of the outgoing waveguides.
This shows the power of the general RSM theory developed here, which will work also for open cavities, where multi-resonance effects dominate.

When ${\bf S}_0 \neq {\bf I}_{N}$, the TMCT analysis above shows that each $\omega_{\rm RZ}$ is an eigenfrequency of
${\bf H}_{\rm RZ} \equiv {\bf H}_{\rm eff} + i {\bf D}^\dagger {\bf S}_0 {\bf F^\dagger} \left( {\bf F }{\bf S}_0 {\bf F}^\dagger \right)^{-1} {\bf F} {\bf D}$.
In the case where ${\bf F }{\bf S}_0 {\bf F}^\dagger$ is not invertible, we can instead apply Eqs.~\eqref{eq:S_in_inverse} \&~\eqref{eq:H_RSM} to ${\bf S S}_0^{-1}$ (instead of directly to ${\bf S}$).
In this case the ``reflectionless'' mode will not be so in the sense defined by ${\bf S}$, but rather by ${\bf SS}_0^{-1}$, which in some cases will actually be a ``transmissionless'' mode of ${\bf S}$.

    \subsection{R-zeros/RSMs as Eigenmodes of the Wave Operator\label{sec:RSM_wave_operator}}

We have introduced two equivalent equations to solve for R-zeros and RSMs: Eq.~\eqref{eq:RSM definition} in terms of a generalized reflection matrix, and Eq.~\eqref{eq:det(A)=0} in terms of an effective wave operator.
Both define nonlinear eigenproblems, and both are applicable to any open system and for each of the $2^N-2$ choices of reflectionless input channels.
While Eq.~\eqref{eq:det(A)=0} has the caveat that it can sometimes also yield BICs or one-sided resonances which are not reflectionless (see discussions at the end of Section~\ref{sec:rigorous RSM}), it is closer to the well-known resonance problem.
In this section we discuss the details of solving for R-zeros through Eq.~\eqref{eq:det(A)=0}.

The distinguishing feature of the wave operator $\hat A_{\rm RZ}$, as defined in Eq.~\eqref{eq:A_RZ}, is the specific frequency-dependent self-energy, formally defined by Eq.~\eqref{eq:RSM self-energy}, that acts on the surface and imposes the boundary condition as outgoing for the channels in $\fbar$ and incoming for the channels in $F$.
Below we describe two implementations, one based on PML and one based on 
an extension of the ``boundary-matching'' conditions associated with purely outgoing or incoming waves~\cite{2005_prog_in_optics,2013_Ambichl_PRX}.

    \subsubsection{PML-based Implementation of Boundary Conditions\label{sec:BC_PML}}

In certain cases, it is possible to solve the R-zero problem without recourse to frequency-dependent boundary conditions, through a mapping to a {\it linear} eigenvalue problem which is readily solved by standard diagonalization (full or partial).
In the absence of dispersion, i.e.,~if the susceptibility is frequency-independent, this mapping can be achieved by using impedance-matched absorbing layers called Perfectly Matched Layers (PMLs)~\cite{1994_Berenger_JCP, 2007_Johnson_PML_note}, which create an effective outgoing boundary through an eigenvalue-independent modification of the bulk wave operator.
PMLs are already commonly used in the calculation of resonances that satisfy purely outgoing boundary conditions~\cite{2018_Lalanne_LPR}.
An incoming boundary condition, while less common, can be similarly implemented with a PML through complex conjugation~\cite{2018_Bonnet-BenDhia_PRSA}.
For example, a left-to-right R-zero is an eigenmode of the wave operator with a conjugated PML on the left and a conventional PML on the right (and vice versa for a right-to-left R-zero); this case was recently described in the context of acoustic waveguides in Ref.~\cite{2018_Bonnet-BenDhia_PRSA}.
The generalization of this to higher dimensions and a greater number of channels is straightforward: simply use an appropriate PML or conjugate PML for the  propagating dimension in each of the asymptotic regions.
Any R-zero problem for which the set of input channels has no spatial overlap with the output channels can be solved by use of this PML-based method; some examples are the six waveguide junction RSM problems shown in Fig.~\ref{fig:octopus}.
Note that PMLs, both conventional and conjugated, must be used with caution, as they introduce additional ``PML modes'' in the eigenvalue spectrum, which need to be identified and discarded~\cite{2018_Lalanne_LPR}.

    \subsubsection{Mode-matching Implementation of Boundary Conditions\label{sec:BC_MM}}

The PML-based method fails for R-zero problems for which the asymptotic input channels are not spatially disjoint from the asymptotic output channels, as described in the introduction.
In these cases, to find a solution we can explicitly match the continuity conditions channel by channel, assigning the appropriate incoming or outgoing condition for that geometry.
We give the explicit, frequency-dependent boundary conditions for three geometries in Appendix~\ref{ap:boundary-matching}, though we cite here the general result for a scalar field $\psi({\bf x})$:
\begin{equation}
    \psi(\bfx)|_{\partial\Omega} = \int_{\partial\Omega} [G^A_F(\bfx,\bfx^\prime)+G^R_{\fbar}(\bfx,\bfx^\prime)] \nabla\psi(\bfx^\prime)\cdot d{\bf S}^\prime,
\end{equation}
where $G^A_F$ is the advanced Green function restricted to the input channels $F$, and $G^R_{\fbar}$ is the retarded green function restricted to the output channels.
The surface integral is over $\partial\Omega$, which separates the scattering region $\Omega$ from the asymptotic regions $\bar\Omega$. 

\begin{figure*}
    \centering
        \includegraphics[width=0.95\textwidth]{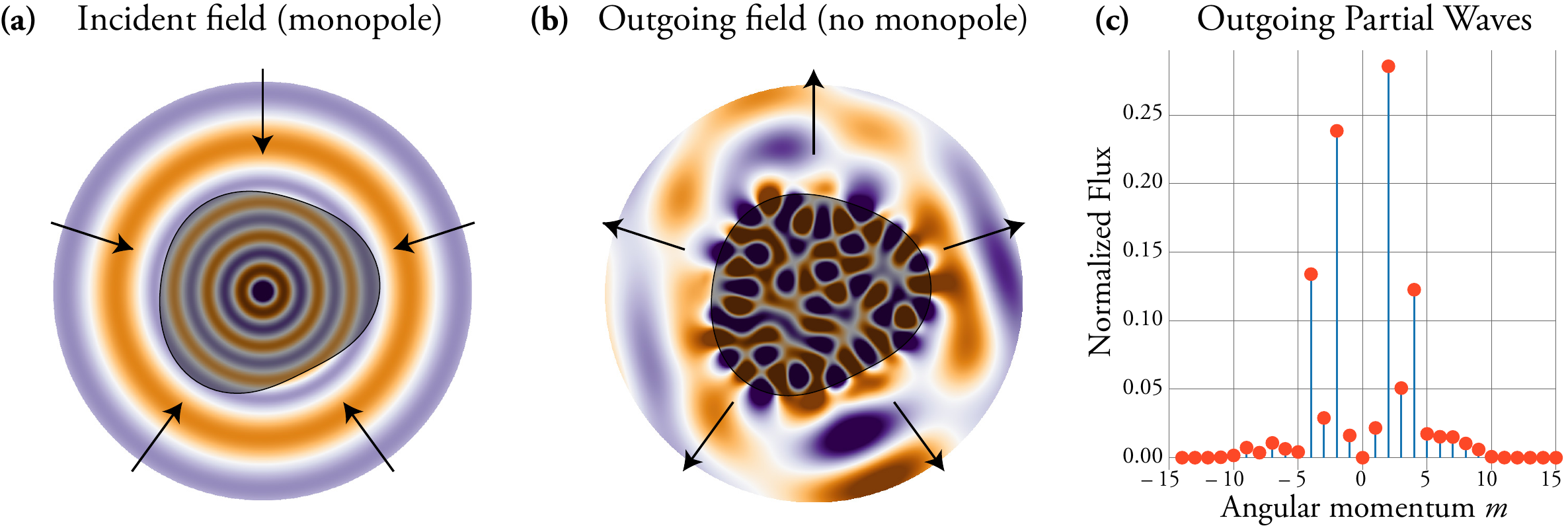}
    \caption{
    (Color) Reflectionless scattering of a monopole scalar wave from a deformed disk dielectric resonator ($n=3)$ with mean radius $\bar R$, shown as shaded region in (a)-(b).
    The disk boundary is $r(\theta)/\bar R=1+x \{.8528\cos(2[\theta-.3127])+.8346\cos(3[\theta-.0079])\}$, with $x=.2768$.
    Monopolar RSM was achieved by tuning the single dimensionless parameter $x$ until a given complex R-zero became real.
    The RSM frequency is at $\omega_0\bar R/c=6.4853$.
    {\bf (a)} The input field only has an $m=0$ component (pure monopole).
    {\bf (b)} The outgoing field contains no $m=0$ component, and is a coherent superposition of many other $m$'s.
    {\bf (c)} The outgoing flux, carried by each angular momentum channel $m$, normalized by the total incident flux. There is no outgoing flux for $m=0$ because the structure was shape-tuned to a monopolar RSM. Note that since we have a purely real index here, the RSM we have found must be bipolar, which implies that if we time-reversed the complicated superposition of outgoing multipole radiation, the incident wavefront would interfere perfectly to generate monopole-only outgoing radiation.
    }
    \label{fig:free_space}
\end{figure*}

In Fig.~\ref{fig:free_space}, we provide an example for which the mode-matching approach must be used: a 2D deformed dielectric resonator in free space, which has been shape-tuned to have an RSM for monopole input at a given complex frequency, $\omega_0$.
The theory here implies that one can perfectly impedance-match a specific superposition of any number of coherent input multipoles to the remaining multipoles.
Thus the scatterer when tuned to RSM acts as a perfectly multipole-transforming antenna.
In the example shown in Fig.~\ref{fig:free_space},
by tuning a single deformation parameter, we were able to find a {\it real} frequency at which the monopole input was reflectionless, so that all of the scattered waves were in higher multipoles.
The R-zero nonlinear eigenvalue problem here was solved using NEP-PACK~\cite{2017_Bezanson_SIAM,2018_Jarlebring_arxiv}.
Through further optimization of the shape one presumably could enhance the scattered output into certain desired outgoing channels. 
Note that the scattering here is not perturbative and the output is not simply determined by single scattering from a particular multipole of the deformation.

The parameter $x$ used to tune to an RSM is the overall strength of deformation, such that $x=0$ yields a circular disk with no deformation.
In the limit that $x\to 0$, continuous rotational symmetry is restored, the angular momentum scattering channels labeled by $m$ decouple from each other, and the scattering matrix is diagonal.
Therefore for $x=0$, the R-zero frequencies are also S-matrix zeros, which are constrained by flux conservation to be in the upper half plane.
On the other hand, as the deformation $x$ is made larger, the $m$ channels become increasingly mixed, so that the single monopole input becomes coupled to more and more multipole outputs.
The effective radiative loss will eventually overtake the effective irradiative gain and push the R-zeros into the lower half-plane.
Therefore by continuity there will be a deformation strength $x_0$ at which the R-zero crosses the real axis, becoming an RSM.

\section{Summary and Conclusions}

This paper presents a significant generalization of the concept of a scattering resonance, which mathematically is defined as a solution of the relevant wave equation in an infinite domain with purely outgoing boundary conditions.
Here we showed that, quite generally, any mixture of partially incoming and (complementary) partially outgoing boundary conditions defines a similar 
resonance phenomenon, which manifests itself as a narrow band, reflectionless scattering process with the same linewidth as the usual resonance,
requiring a specific coherent input state determined from the reflectionless scattering mode (RSM). 
The framework given here applies to all linear classical wave scattering, and to quantum scattering as well. 
Without tuning of the parameters of the wave operator, such reflectionless states (R-zeros) exist at complex frequency, and do not in general appear as real frequency steady-state solutions. 
In the case of ${\cal P}$ and ${\cal T}$  symmetric (lossless) scatterers they do appear as the bidirectional unit transmission resonances familiar from elementary textbooks examples. 
For the case of $\mathcal{PT}$ symmetric scatterers, steady-state solutions also exist generically, but they are {\it unidirectional} and disappear above a spontaneous $\mathcal{PT}$ symmetry-breaking transition. 
Both the $\mathcal{PT}$ and $\mathcal{P,T}$ symmetry cases can lose their RSMs by passing through an exceptional point (EP), beyond which the solutions leave the real-frequency axis. 
These represent a new type of EP in terms of their physical manifestation, and unlike the widely-studied resonant EPs, they can occur on the real axis without generating a self-oscillation instability.

Our results can be most naturally applied to classical electromagnetic scattering and acoustic scattering. 
In this context they can be regarded as defining necessary and sufficient conditions for the existence of perfectly impedance-matched solutions for a given scattering geometry. 
Roughly speaking we have shown that such solutions do not exist generically for the wave equations of interest in physics, barring special symmetries.
But with a single continuous tuning parameter those impedance-matched states corresponding to the RSM boundary conditions can be engineered 
to exist. 
Other desirable states cannot be engineered by tuning a single parameter; e.g. similar to~\cite{2017_Su_ACSPh}, in a three-mode waveguide junction a state with input in waveguide one, which is reflectionless and only scatters into waveguide two and not into waveguide three, does not correspond to the RSM boundary condition. 
It cannot in general be achieved with single parameter tuning and is not guaranteed to exist if more parameters are tuned.
Nonetheless, if such a state is desired one could search for it by first finding the RSM with one channel in and two out, and then optimizing further parameters so as to minimize the scattering into waveguide three.
Hence we are optimistic that our RSM theory can provide a powerful tool for design in photonics and acoustics.

Since the theory of RSMs can determine a perfectly impedance-matched steady-state of linear Maxwell electrodynamics, it will also determine an impedance matched state for quantum electrodynamics, for which all moments of the reflected flux will vanish at $\omega_{\rm RZ}$.
Quantum fluctuations will arise only due to the finite linewidth of the input field \cite{2013_Chong_PRA}.
Therefore the RSM concept can be of interest in quantum optics too.
We note that since absorption-gain tuning is not necessary to create RSMs (although it is necessary to make them unidirectional), it can be applied to quantum systems without reservoir tuning.

\begin{acknowledgments}
W.R.S. and A.D.S. acknowledge the support of the NSF CMMT program under grant DMR-1743235.
\end{acknowledgments}

\appendix

\section{Derivation of determinant relations}

In this appendix we derive \eqref{eq:det(S)} for an $N$-channel S-matrix, and and~\eqref{eq:det(R)} for an $N_{\rm in}$-input generalized reflection matrix ${\bf R}_{\rm in}$, using the two identites
\begin{gather}
     ({\bf A+BC})^{-1}{\bf B}={\bf A}^{-1}{\bf B}({\bf I}_M+{\bf CA}^{-1}{\bf B})^{-1} \label{eq:push-through},\\
     \det({\bf I}_N-{\bf BC}) = \det({\bf I}_M-{\bf CB}) \label{eq:Sylvester},
\end{gather}
for invertible ${\bf A}\in{\mathbb C}^{N\times N}$, and arbitrary ${\bf B}\in{\mathbb C}^{N\times M}$, ${\bf C}\in{\mathbb C}^{M\times N}$, which we now derive.

The first identity is a generalization of the ``push-through identity''~\cite{Bernstein_Matrix_book}
\begin{equation}
    \label{eq:baby push_through}
    ({\bf I}_N+{\bf BD})^{-1}{\bf B}={\bf B}({\bf I}_M+{\bf DB})^{-1},
\end{equation}
with ${\bf D}\in{\mathbb C}^{M\times N}$, and ${\bf B}$ as before.
It is named for its action on ${\bf B}$ relative to the inverse, and follows trivially from noting that ${\bf B}({\bf I}_M+{\bf DB})=({\bf I}_N+{\bf BD}){\bf B}$.
This can be generalized for ${\bf A,B,C}$, with ${\bf A}\in{\mathbb C}^{N\times N}$ invertible by starting with $({\bf A}+{\bf BC})^{-1} = {\bf A}^{-1}({\bf I}_N+{\bf BCA}^{-1})^{-1}$.
Applying Eq.~\eqref{eq:baby push_through}, with ${\bf D} = {\bf C}{\bf A}^{-1}$, we arrive at
\begin{equation}
\label{eq:genearlized push-through}
    ({\bf A}+{\bf BC})^{-1}{\bf B}={\bf A}^{-1}{\bf B}({\bf I}_M+{\bf CA}^{-1}{\bf B})^{-1},
\end{equation}
which proves the first identity.

The second identity can be derived as a special case of Schur's determinant formula~\cite{Handbook_of_LA}
\begin{equation}
    \label{eq:Schur}
    \det {\bf A} \det({\bf D} - {\bf C}{\bf A}^{-1} {\bf B}) = \det {\bf D} \det({\bf A} - {\bf B}{\bf D}^{-1} {\bf C})
\end{equation}
where ${\bf A,B,C}$ are defined as before, and ${\bf D}\in{\mathbb C}^{M\times M}$ is invertable.
The judicious choice ${\bf A} = {\bf I}_N$, ${\bf D}={\bf I}_M$ gives
\begin{equation}
    \det ({\bf I}_N-{\bf BC}) = \det ({\bf I}_M-{\bf CB}),
\end{equation}
which is what we wanted to show.

    \subsection{Evaluating $\det {\bf S}(\omega)$ \label{ap:detailed_derivation_S}}

Applying Eq.~\eqref{eq:push-through} to $({\bf A}_0^\prime-{\boldsymbol \Delta}+i{\boldsymbol \Gamma})^{-1} {\bf W}_p$, recalling that ${\boldsymbol \Gamma}=\pi {\bf W}_p{\bf W}_p^\dagger$, yields
\begin{equation}
    ({\bf A}_0^\prime-{\boldsymbol \Delta} + i{\boldsymbol \Gamma})^{-1} {\bf W}_p = {\bf G}_0^{\prime\prime} {\bf W}_p({\bf I}_N+i\pi {\bf W}_p^\dagger {\bf G}^{\prime\prime}_0 {\bf W}_p)^{-1},
\end{equation}
where ${\bf G}_0^{\prime\prime}=({\bf A}_0^\prime-{\boldsymbol \Delta})^{-1}$.
Using this in the full expression for the S-matrix (\ref{eq: Heidelberg},~\ref{eq:Aeff_and_Sigma}), and factoring out $({\bf I}_N+i\pi {\bf W}_p^\dagger {\bf G}_0^{\prime\prime} {\bf W}_p)^{-1}$:
\begin{equation}
    {\bf S} = ({\bf I}_N-i\pi {\bf W}_p^\dagger {\bf G}_0^{\prime\prime} {\bf W}_p)/({\bf I}_N+i\pi {\bf W}_p^\dagger {\bf G}_0^{\prime\prime} {\bf W}_p).
\end{equation}
Taking the determinant and applying Eq.~\eqref{eq:Sylvester} to the numerator and denominator, we have
\begin{align}
    \det {\bf S} &= \frac{\det ({\bf I} - i\pi {\bf G}_0^{\prime\prime} {\bf W}_p{\bf W}_p^\dagger)}{\det({\bf I} + i\pi {\bf G}_0^{\prime\prime} {\bf W}_p{\bf W}_p^\dagger)}\nonumber\\
    & = \frac{\det ({\bf A}_0^\prime - {\boldsymbol \Delta} - i{\boldsymbol \Gamma})}{\det({\bf A}_0^\prime - {\boldsymbol \Delta} + i{\boldsymbol \Gamma})},
\end{align}
where ${\bf I}$ is the identity on the (infinite-dimensional) closed-cavity Hilbert space.
This proves Eq.~\eqref{eq:det(S)}.

    \subsection{Evaluating $\det {\bf R}_{\rm in}(\omega)$\label{ap:detailed_derivation_Rin}}

We proceed as we did with the S-matrix, but starting with ${\bf R}_{\rm in}$ from~\eqref{eq:Rin first}, where $F$ defines the inputs: push ${\bf W}_F$ through ${\bf G}_{\rm eff}$, factor out a common inverse, take the determinant and apply~\eqref{eq:Sylvester}.

Writing ${\bf G}_{\rm eff}$ in~\eqref{eq:Rin first} in terms of ${\bf W}_F$ and ${\bf W}_{\fbar}$, and using the identity~\eqref{eq:push-through} yields
\begin{equation}
    (\bar{\bf A}_0+i{\boldsymbol \Gamma}_F)^{-1} {\bf W}_F = \bar{\bf A}_0^{-1} {\bf W}_F({\bf I}_{N_{\rm in}}+i\pi {\bf W}_F^\dagger \bar{\bf A}_0^{-1} {\bf W}_F)^{-1},
\end{equation}
where $\bar{\bf A}_0 \equiv {\bf A}_0^\prime-{\boldsymbol \Delta} + i{\boldsymbol \Gamma}_{\fbar}$.
See~\eqref{eq:Aeff_and_Sigma} and~\eqref{eq:RSM self-energy} for the definitions of ${\boldsymbol \Delta}$, ${\boldsymbol \Gamma}_{F,\fbar}$, and ${\bf W}_{F,\fbar}$.

Plugging this into~\eqref{eq:Rin first} and factoring out $({\bf I}_{N_{\rm in}}+i\pi {\bf W}_F^\dagger \bar{\bf A}_0^{-1} {\bf W}_F)^{-1}$ gives a K-matrix representation for ${\bf R}_{\rm in}$:
\begin{equation}
{\bf R}_{\rm in} = ({\bf I}_{N_{\rm in}} - i \pi {\bf W}_F^\dagger \bar{\bf A}_0^{-1} {\bf W}_F)/({\bf I}_{N_{\rm in}} + i \pi {\bf W}_F^\dagger \bar{\bf A}_0^{-1} {\bf W}_F).
\end{equation}

Taking the determinant, using the identity~\eqref{eq:Sylvester}, and multiplying the numerator and denominator by $\det \bar{\bf A}_0$ results in
\begin{align}
    \det {\bf R}_{\rm in} &= \frac{\det ({\bf I}-i\pi \bar {\bf A}_0^{-1}{\bf W}_F{\bf W}_F^\dagger)}{\det ({\bf I}+i\pi \bar {\bf A}_0^{-1}{\bf W}_F{\bf W}_F^\dagger)}\\
    &=\frac{\det ({\bf A}^\prime-{\boldsymbol \Delta}+i{\boldsymbol \Gamma}_{\fbar}-i{\boldsymbol \Gamma}_F)}{\det ({\bf A}_0^\prime-{\boldsymbol \Delta}+i{\boldsymbol \Gamma}_{\fbar}+i{\boldsymbol \Gamma}_F)}.
\end{align}
Finally, using the identities for the filters ${\bf F}$ and $\bar{\bf F} $~\eqref{eq:F identites}, the denominator can be simplified by ${\boldsymbol \Gamma}_F + {\boldsymbol \Gamma}_{\fbar}={\boldsymbol \Gamma}$.
This proves Eq.~\eqref{eq:det(R)}.

\section{Explicit forms of boundary-matching \label{ap:boundary-matching}}

(1) In 1D, with the scattering region contained entirely in $|x|<a$, the RSM bc's are
\begin{align}
    \label{eq:d=1 bc}
    \text{left-incident: }\psi(\pm a)&=+\partial_x \psi(\pm a)/i(\omega/c)\\
    \text{right-incident: }\psi(\pm a) &= -\partial_x \psi(\pm a)/i(\omega/c).
\end{align}

(2) For a metallic waveguide with transverse width $t$ in 2D, and with the scattering region contained entirely in $|x|<a$, the RSM bc's are
\begin{equation}
    \psi(\pm a,y) = \mp \int dy^\prime K_{F_\pm}(y,y^\prime)\ \partial_x\psi(\pm a,y^\prime)
\end{equation}
where $F_-$ is the set of input channels for the left lead, and $F_+$ for the right. The kernel $K_F$ is
\begin{equation}
    K_F(y,y^\prime) = \sum_{m\in F}g^-_m(y,y^\prime) + \sum_{m\notin F}g^+_m(y,y^\prime),
\end{equation}
where
\begin{equation}
    g^\pm_m(y,y^\prime) = \pm\frac{1}{i\beta_m^\pm(\omega)}\sin\Big(\frac{m \pi y}{t}\Big) \sin\Big(\frac{m \pi y^\prime}{t}\Big).
\end{equation}
For real $\omega$, the propagation constant is
\begin{align}
    \beta^\pm_m (\omega)&=\sqrt{\left(\frac{\omega}{c}\right)^2-\left(\frac{m\pi}{t}\right)^2\pm i0^+}.
\end{align}
The square-root branch cut is the conventional one along the negative real axis, such that $\sqrt{-1\pm i0^+}=\pm i$.
It is worth noting that the contribution of each propagating mode to $K_F$ has a sign which depends on whether the mode is designated as input or output, while the non-propagating modes all contribute with the same sign, regardless of the choice of $F$.
For complex $\omega$, the construction of the propagation constant $\beta_m$ is more involved, and beyond the scope of this paper~\footnote{
The propagation constant $\beta_m(\omega)$ for each mode has a sign ambiguity. One may try to resolve it for propagating modes by considering which sign corresponds to the correct sign of ${\rm Re}\{\beta_m$\}, as determined by $F$, and for evanescent modes by which one gives exponential decay far from the scattering region.
However, the distinction between propagating and evanescent modes is itself ambiguous for complex $\omega$, and must be determined by the asymptotic behavior of each mode when continued to the real axis.}.

(3) A finite scatterer in 2D free-space, contained entirely within a radius $R$, has asymptotic channels specified by angular momentum $m$. The R-zero boundary condition is
\begin{equation}
    \psi(R,\phi) = \sum_{m} \frac{e^{im(\phi-\phi^\prime)}}{2\pi c^F_m(kR)} \partial_r \psi(R,\phi^\prime),
\end{equation}
where 
\begin{equation}
    c^F_m(x) = 
    \begin{cases}
    \partial_R \ln H^{(2)}_m(x),&m\in F\\
    \partial_R \ln H^{(1)}_m(x),&m\notin F.
    \end{cases}
    \label{eq:d=2 bc}
\end{equation}
$H_m^{(1,2)}(x)$ are the Hankel functions of the first and second kind (outgoing and incoming, respectively) of order $m$~\cite{1972_Abramowitz_book}.

All of the boundary conditions delineated above (Eqs.~\eqref{eq:d=1 bc}--\eqref{eq:d=2 bc}) are specific instances of a general formula which relates the function at the boundary $\partial\Omega$ to its normal derivative via the appropriate Green function, which is advanced in the input channels $F$ and retarded in the output channels $\fbar$:
\begin{equation}
    \psi(\bfx)|_{\partial\Omega} = \int_{\partial\Omega} [G^A_F(\bfx,\bfx^\prime)+G^R_{\fbar}(\bfx,\bfx^\prime)] \nabla\psi(\bfx^\prime)\cdot d{\bf S}^\prime.
\end{equation}
The derivation of this takes us too far from the main thread of this work, and will be discussed in a future publication.

\bibliography{references}

\end{document}